\documentclass[12pt,a4paper]{article}
\usepackage[usenames,dvipsnames]{pstricks} 
\usepackage{afterpage}
\usepackage{epsfig}
\usepackage{pst-grad} 
\usepackage{pst-plot} 
\usepackage[a4paper]{geometry}
\usepackage{float}
\usepackage[stable]{footmisc}
\usepackage[utf8x]{inputenc}
\usepackage{a4wide}
\usepackage{amsmath,amssymb,amstext,amsthm,amssymb}
\usepackage{amsfonts}
\usepackage{framed}
\usepackage{graphicx}
\usepackage{color}
\usepackage{bbold}
\usepackage{bbm}
\usepackage{rotating} 
\usepackage{slashed} 
\usepackage{cancel} 
\usepackage{braket} 
\usepackage{amstext}
\usepackage{array}
\usepackage{setspace}
\usepackage{hyperref}
\usepackage{overpic}
\usepackage{bbm}
\usepackage{cite}
\usepackage{lipsum}
\usepackage{dsfont}

\newcommand{\del}[0]{\partial}

\let\baraccent=\=
\renewcommand{\=}[1]{\stackrel{#1}{=}}
\newcommand{\id}[0]{\mathds{1}}

\DeclareSymbolFontAlphabet{\mathbb}{AMSb}

\begin{document}

\pagestyle{plain}

\makeatletter
\@addtoreset{equation}{section}
\makeatother
\renewcommand{\theequation}{\thesection.\arabic{equation}}
\pagestyle{empty}
\vspace{0.5cm}

\begin{center}
	
	{\LARGE \bf{Orientifolding Kreuzer-Skarke} \\[10mm]}
\end{center}

\begin{center}
	\scalebox{0.95}[0.95]{{\fontsize{14}{30}\selectfont Jakob Moritz}}
\end{center}

\begin{center}
	\vspace{0.25 cm}
	
		\textsl{Department of Physics, Cornell University, Ithaca, NY 14853, USA}\\

	\vspace{1cm}
	\normalsize{\bf Abstract} \\[8mm]

\end{center}

We develop tools that allow the systematic enumeration of inequivalent holomorphic orientifolds of Calabi-Yau hypersurfaces in toric fourfolds of arbitrary Hodge numbers. As examples, we construct an orientifold of the Calabi-Yau hypersurface with largest known Hodge number $h^{1,1}=491$, as well as an orientifold of a Calabi-Yau hypersurface with $h^{1,1}=243$ that yields a large orientifold-odd Hodge number $h^{1,1}_-=120$.

\begin{center}
	\begin{minipage}[h]{15.0cm}

	\end{minipage}
\end{center}
\newpage
\setcounter{page}{1}
\pagestyle{plain}
\renewcommand{\thefootnote}{\arabic{footnote}}
\setcounter{footnote}{0}
%
%
\tableofcontents
\newpage

	\section{Introduction}
	In order to understand how the microscopic features of string theory give rise to realistic particle physics and cosmology at long distances one must understand, and chart out, the set of four dimensional low energy effective field theories that arise as infrared limits of string theory compactifications. The best understood such theories preserve eight or more supercharges in four dimensions and arise via compactification of type II string theories on Calabi-Yau (CY) threefolds, or via heterotic/type I string theory on K3$\times T^2$. Breaking supersymmetry weakly down to four or zero supercharges via orientifold projections, fluxes and spacetime filling D-branes gives rise to the famous landscape of (semi-)realistic string theory vacua \cite{Giddings:2001yu,Kachru:2003aw,Ashok:2003gk,Denef:2004ze}.
	
	The UV data that encodes any EFT in this landscape is a choice of CY threefold, an orientifold involution thereof, and a consistent configuration of D-branes and fluxes. Thus, a crucial part of understanding the string landscape involves studying CY threefolds and their orientifold involutions: this data controls the spectrum of light chiral and vector multiplets, in particular the spectrum of light axion like particles. While a wealth of tools exists that allows the systematic construction of CY threefolds, e.g. as hypersurfaces in toric varieties \cite{Batyrev:1993oya,Kreuzer:2000xy}, a similarly systematic study of their orientifold involutions is so far lacking.\footnote{See however \cite{Gao:2013pra,Carta:2022web,Crino:2022zjk,Shukla:2022dhz} for results on orientifolds of hypersurfaces in toric varieties with small Hodge number $h^{1,1}$.} Exceptions include a systematic study of orientifolds of complete intersection CY threefolds in products of projective spaces \cite{Carta:2020ohw}.
	
	In this paper we will remedy this: we explain how to efficiently construct the set of holomorphic orientifold involutions associated to any of the  $473,800,776$  (birational equivalence classes of) CY threefolds in the Kreuzer-Skarke database. We then demonstrate our ability to treat even the CY threefolds with largest Hodge numbers by detailing examples of orientifolds of CY threefold hypersurfaces with Hodge numbers $h^{1,1}=491$, and $h^{1,1}=243$, a regime of Hodge numbers that had so far been out of reach. In particular we will display an example with orientifold splitting of the Hodge number $h^{1,1}=243\rightarrow (h^{1,1}_+,h^{1,1}_-)=(123,120)$, the first example known to us that gives rise to a large number of two-form axions.\footnote{Phenomenological aspects of two-form axions have been studied e.g. in \cite{McAllister:2008hb,Hebecker:2015tzo,Hebecker:2018yxs,Carta:2021uwv,Cicoli:2021gss,Cicoli:2021tzt,Carta:2022web}.}
	
	This paper is structured as follows. We start with \S\ref{sec:orientifolds_background_material}, reviewing well known results about the effective field theories from CY orientifolds \cite{Grimm:2004uq}, and a brief summary of the construction of CY hypersurfaces in toric varieties as in \cite{Batyrev:1993oya,Kreuzer:2000xy}. The main part of this paper is \S\ref{sec:orientifolding_KS} in which we explain our results. We display our examples in \S\ref{sec:examples} and conclude in \S\ref{sec:conclusions}. 
	
	\section{Orientifolds of Calabi-Yau threefolds}\label{sec:orientifolds_background_material}
	In this section we review well known facts about Calabi-Yau orientifolds of type IIB string theory, following \cite{Grimm:2004uq}.  The massless bosonic spectrum of ten-dimensional type IIB string theory consists of the metric $g_{MN}$, the two form $B_2$, and the dilaton $g_s\equiv e^{\phi}$ from the NS-NS sector, as well as the $p$-form gauge fields $C_p$, with $p=0,2,4$, from the RR sector. The RR four-form has self-dual field strength.
	
	Compactifying on a Calabi-Yau threefold preserves $\mathcal{N}=2$ supersymmetry in four dimensions, and the massless spectrum arises from the Kaluza-Klein zero-modes of the ten-dimensional fields, as analyzed e.g. in \cite{Bohm:1999uk}. As usual we denote the Hodge numbers of $X$ by 
	\begin{equation}
		h^{p,q}(X):=\text{dim}_{\mathbb{R}}H^{p,q}(X,\mathbb{R})\, .
	\end{equation}
	Furtherore, we denote by $[H_a]$, $a=1,\ldots,h^{1,1}(X)$ a basis of $H^2(X,\mathbb{Z})$, and by $[\mathcal{C}^a]$ the dual basis of $H^4(X,\mathbb{Z})$, and we let $([\alpha_I],[\beta^I])$, $I=0,\ldots,h^{2,1}(X)$, be a symplectic basis of $H^3(X,\mathbb{Z})$. In terms of these we define the periods
	\begin{equation}
		t^a:=\int_X J_X\wedge [\mathcal{C}^a]\, ,\quad Z^I:=\int_X \Omega\wedge [\beta^I]\, ,\quad \mathcal{F}_I:=\int_X \Omega\wedge [\alpha_I]\,  ,
	\end{equation}
	where $J_X$ is the K\"ahler form on $X$ and $\Omega$ is the holomorphic threeform. The $t^a$ are local coordinates on K\"ahler moduli space, and the $Z^I$ are local projective coordinates on complex structure moduli space. Away from $Z^0=0$ we can define local affine coordinates $z^i:=Z^i/Z^0$, $i=1,\ldots, h^{2,1}(X)$.
	
	Moreover we define the (pseudo-)scalars
	\begin{equation}\label{eq:N=2axions}
		\xi_a:=\int_X C_4\wedge [H_a]\, ,\quad c^a:=\int_X C_2\wedge [\mathcal{C}^a]\, ,\quad b^a:=\int_X B_2\wedge [\mathcal{C}^a]\, ,
	\end{equation}
	and the $U(1)$ vectorfields
	\begin{equation}\label{eq:N=1vectors}
		A^I:=\int_X C_4\wedge [\beta^I]\, .
	\end{equation}
	The four dimensional effective supergravity theory is furnished by the gravity multiplet containing as bosonic components the four dimensional metric $g_{\mu\nu}$ and one linear combination of the $A^I$, as well as $h^{2,1}$ vector multiplets containing the complex structure moduli $z^i$ and the remaining linear combinations of the $A^I$. Finally there are $h^{1,1}$ hypermultiplets containing the K\"ahler moduli $t^a$, and the pseudo-scalars $(\xi_a,c^a,b^a)$, and a further \emph{universal} hypermultiplet containing the zero modes of $e^\phi$ and $C_0$, as well as the four-dimensional electric-magnetic duals of the two forms $C_2$ and $B_2$.
	
	Now let $\mathcal{I}$ be a holomorphic involution of $X$ with trivial induced action on the K\"ahler class $[J]\mapsto [J]$. Its induced action on cohomology defines the orientifold-even and odd Hodge numbers
	\begin{equation}
		h^{p,q}_{\pm}:=\text{dim}_{\mathbb{R}}\left[(\id\pm \mathcal{I}^*) H^{p,q}(X,\mathbb{R})\right]\, ,
	\end{equation}
	where $\mathcal{I}^*$ is the induced action of $\mathcal{I}$ on cohomology. 
	
	Involutions $\mathcal{I}$ come in two different classes, determined by the induced action on the holomorphic threeform \cite{Dabholkar:1996pc,Sen:1996vd,Acharya:2002ag,Brunner:2003zm,Brunner:2004zd}:
	\begin{align}\label{eq:orientifold_types}
		&\text{O3/O7}:\quad \mathcal{I}^*\, : \quad [\Omega]\mapsto -[\Omega]\, ,\\
		&\text{O5/O9}:\quad \mathcal{I}^*\, : \quad [\Omega]\mapsto +[\Omega]\, .
	\end{align}
	Involutions of O3/O7 type have a fixed point set of odd co-dimension, a union of points and divisors, while involutions of O5/O9 type have fixed points at even co-dimension.
	
	First, let $\mathcal{I}$ be of O3/O7 type. Then, truncating the string spectrum to the part invariant under $(-1)^{F_L}\circ \omega \circ \mathcal{I}$, where $F_L$ is left moving spacetime fermion number, and $\omega$ is worldsheet parity, produces a Calabi-Yau orientifold vacuum preserving $\mathcal{N}=1$ supersymmetry in four dimensions, hosting O3 planes on the isolated fixed points of $\mathcal{I}$, and O7 planes along the co-dimension one components of the fixed point set. For suitable D-brane configurations filling the $4d$ spacetime and orientifold odd threeform field strength backgrounds $H_3=dB_2$ and $F_3=dC_2$ one obtains consistent (i.e. free of infra-red divergences) physical theories: letting $\mathcal{F}_{\mathcal{I}}$ be the fixed point set in $X$ and $[\text{O7}]\in H^2(X,\mathbb{Z})$ the class of the co-dimension one component of $\mathcal{F}_{\mathcal{I}}$ one must have
	\begin{equation}
		[D7]=8[\text{O7}]\, ,\quad Q_{D3}:=\frac{\chi(\mathcal{F}_I)}{4}=\frac{1}{2}\int_X H_3\wedge F_3+N_{D3}\, ,
	\end{equation}
	where $\chi$ is the topological Euler characteristic, $[D7]$ is the four-cycle class of the D7 brane, and $N_{D3}$ is the number of mobile D3 branes at points in $X$. Here, a D3 brane stuck on an orientifold plane contributes with $+\frac{1}{2}$ to $N_{D3}$ while a fully mobile D3 brane contributes with $+1$.  In general the divisor class $8[\text{O7}]$ has normal bundle moduli, in which case there is a continuous moduli space of consistent D7 brane configurations, see \cite{Braun:2008ua,Collinucci:2008pf} for details. Wrapping a stack of eight D7 branes on each irreducible component at co-dimension one of the fixed point set yields a non-abelian gauge algebra $\mathfrak{so}(8)^{n_{\text{O7}}}$, where $n_{\text{O7}}$ is the number of O7 planes.
	
	The massless closed string spectrum of the physical theory is obtained by projecting the $\mathcal{N}=2$ multiplets of the Calabi-Yau compactification to $\mathcal{N}=1$ multiplets: the $\mathcal{N}=2$ gravity multiplet projects to the $\mathcal{N}=1$ gravity multiplet. The $h^{2,1}$ $\mathcal{N}=2$ vector multiplets project to $h^{2,1}_-$ chiral multiplets furnished by the complex structure deformations that preserve \eqref{eq:orientifold_types}, and $h^{2,1}_+$ $\mathcal{N}=1$ vector multiplets corresponding to the symmetric $(2,1)$ forms in \eqref{eq:N=1vectors}. The $h^{1,1}$ non-universal hypermultiplets project to $h^{1,1}_+$ chiral mutiplets containing the deformations of the K\"ahler class that preserve the invariance under $\mathcal{I}^*$, and the linear combinations of four form axions $\xi_a$ corresponding to even four forms, as well as $h^{1,1}_-$ chiral multiplets containing the linear combinations of two form axions $\mathcal{G}^a:=c^a-\tau b^a$ associated with the odd two forms. The universal hypermultiplet projects to the \emph{axio-dilaton} $\tau:=C_0+ie^{-\phi}$.
	
	The classical superpotential is the Gukov-Vafa-Witten (GVW) flux superpotential \cite{Gukov:1999ya,Giddings:2001yu}
	\begin{equation}
		W_{\text{flux}}=\int_X (F_3-\tau H_3)\wedge \Omega\, ,
	\end{equation}
	and receives corrections from gaugino condensation effects on seven-branes, as well as from D3 instantons with gauge bundles wrapping holomorphic divisors \cite{Witten:1996bn,Bianchi:2011qh,Bianchi:2012kt,Bianchi:2012pn,Alexandrov:2022mmy,Gendler:2022qof,Kim:2022uni,Kim:2023cbh,Jefferson:2022ssj}, as well as from D(-1) instantons.\footnote{Note that D(-1) instantons do not contribute at the point in moduli space where a stack of eight D7 branes is wrapped on top of each O7 plane \cite{Kim:2022jvv}.}
	
	In addition to the closed string multiplets obtained by projecting the $\mathcal{N}=2$ multiplets to $\mathcal{N}=1$ multiplets one also gets open-string multiplets localized on the background D-branes.
	
	In contrast, for $\mathcal{I}$ of O5/O9 type the string spectrum can be truncated by keeping the modes invariant under $\omega\circ \mathcal{I}$. In this case the involution is either trivial, corresponding to a Calabi-Yau compactification of type I string theory, or has fixed points at complex co-dimension two, corresponding to an O5 orientifold of type IIB string theory. Again, the $\mathcal{N}=2$ gravity multiplet projects to the $\mathcal{N}=1$ gravity multiplet. The $h^{2,1}$ $\mathcal{N}=2$ vector multiplets project to $h^{2,1}_+$ chiral multiplets corresponding to the complex structure deformations preserving \eqref{eq:orientifold_types}, and $h^{2,1}_-$ $\mathcal{N}=1$ vector multiplets containing the $\xi_a$ corresponding to the odd four forms. The $h^{1,1}$ universal hypermultiplets project to $h^{1,1}_+$ chiral multiplets containing the deformations of the K\"ahler class preserving $\mathcal{I}^*$-invariance and the linear combinations of two form axions $c^a$ corresponding to even two-forms, as well as $h^{1,1}_-$ chiral multiplets containing the $(\xi_a,b^a)$ associated with odd two and four forms.
	
	At tree level in the $\alpha'$ and string loop expansion, the K\"ahler potential and the holomorphic coordinates on moduli space are well-known, and we refer the reader to the original references \cite{Grimm:2004uq}. More generally, for O3/O7 orientifolds the full K\"ahler potential at closed string tree level is known to all orders in $\alpha'$ both perturbatively \cite{Becker:2002nn} as well as non-perturbatively (see e.g. \cite{Demirtas:2021nlu}).
	
	\section{Calabi-Yau hypersurfaces from toric varieties}\label{sec:CYhypersurfaces}
	In this section we recall a few basic facts (see e.g. \cite{Hori:2003ic}) about toric fourfolds constructed from $4d$ reflexive polytopes and their CY hypersurfaces, as pioneered by Batyrev \cite{Batyrev:1993oya}.
	
	Let $N\simeq \mathbb{Z}^4$ be a 4d lattice, let $N_{\mathbb{R}}:=\mathbb{R}\otimes N\simeq \mathbb{R}^4$, and let $M$ be the lattice dual to $N$ and similarly $M_{\mathbb{R}}:=M\otimes \mathbb{R}\simeq \mathbb{R}^4$. Furthermore, let $\Delta^\circ\subset N_{\mathbb{R}}$ be a $4d$ reflexive polytope and $\mathcal{T}$ a fine, regular, and star triangulation (FRST) of $\Delta^\circ \cap N$ ignoring the points interior to facets of $\Delta^\circ\cap N$. Let $V$ be the toric variety with toric fan $\Sigma_{\mathcal{T}}$ defined by the set of cones over the simplices of $\mathcal{T}$. To each point $p\in \mathcal{P}^\circ:= (\Delta^\circ\cap N)\backslash\{0\}$ we associate a toric coordinate $x_p$ and a toric divisor $\hat{D}_p:=\{x_p=0\}$. We can over-parameterize $V$ by \textit{homogeneous coordinates} $\{x_p,\ldots,x_{n+4}\}\in \mathbb{C}^{n+4}\backslash Z$ where $n:=h^{1,1}(V)=|\mathcal{P}^\circ|-4$, and the exclusion set $Z$ is the union over all subloci $\{x_{p_1}=x_{p_2}=...=x_{p_k}=0\}$ of $\mathbb{C}^{n+4}$ with $\{p_1,...,p_k\}$ any subset of $\mathcal{P}^\circ$ s.t. the cone over those points is \textit{not} a cone in the toric fan $\Sigma$. 
	
	We have
	\begin{equation}\label{eq:toric_isomorphism}
		V\simeq (\mathbb{C}^{n+4}\backslash Z)/G\, ,
	\end{equation}
	and the group of toric scaling relations $G$ is defined as
	\begin{equation}
		G:=\left\{\eta \in \left(\mathbb{C}/\mathbb{Z}\right)^{n+4}: \sum_{p\in \mathcal{P}^0}\eta^p p\in N\right\}\, ,
	\end{equation}
	and acts on $\mathbb{C}^{n+1}\backslash Z$ via $\eta:\, x_p\mapsto e^{2\pi i \eta^p}x_p$. 
	
	An algebraic torus $T:=(N\otimes \mathbb{C})/N\simeq (\mathbb{C}^*)^4$ action on $V$ can be defined via the following action of a representative $t\in N\otimes \mathbb{C}$ on representatives in $\mathbb{C}^{n+4}\backslash Z$,
	\begin{equation}\label{eq:algebraictorus}
		\phi_{[t]}:\, x_p\mapsto e^{2\pi i \eta_t^p}x_p\, ,
	\end{equation}
	with $\eta_t$ obtained by decomposing $t=\sum_{p\in \mathcal{P}^0} \eta_t^p p$. Choosing a different representative $t'=t\mod N$, or a different decomposition changes the action of $[t]$ on the homogeneous coordinates at most by an element of $G$, so \eqref{eq:algebraictorus} is well-defined. We have
	\begin{equation}
		\phi_{[t]}\circ \phi_{[t']}=\phi_{[t+t']}\, .
	\end{equation}
	We denote by
	\begin{equation}
		\hat{D}_p:=\{x_p=0\}\subset V\ ,
	\end{equation}
	for $p\in \mathcal{P}^\circ$,
	the toric divisors of $V$ corresponding to the edges of the toric fan $\Sigma_{\mathcal{T}}$. Their cohomology classes span the lattice $H^2(V,\mathbb{Z})=H^{1,1}(V,\mathbb{Z})$. In terms of these, we may write a general K\"ahler form on $V$ as
	\begin{equation}\label{eq:kahler_V}
		J_V=\sum_{p\in \mathcal{P}^\circ}\mathtt{t}_p [\hat{D}_p]\, .
	\end{equation}
	
	
	Next, we define  $\Delta$ as the polar dual of $\Delta^\circ$,
	\begin{equation}
		\Delta:=\{q\in M_{\mathbb{R}}| \langle p,q \rangle \geq -1 \, ,\forall p\in \Delta^\circ \}\, .
	\end{equation}
	A basis of global sections of the anti-canonical line bundle $K_V^{-1}$ of $V$ is given by the monomials
	\begin{equation}
		s_q:=\prod_{p\in \mathcal{P}^\circ}x_p^{\langle p,q \rangle+1}\, .
	\end{equation}
	We let $X$ be a (suitably generic) anti-canonical hypersurface which is Calabi-Yau (CY), i.e. 
	\begin{equation}\label{eq:CY_hypersurface}
		X:=\{f=0\}\, ,\quad f:=\sum_{q\in \Delta\cap M}\psi_q s_q\, ,
	\end{equation}
	with complex parameters $\psi_q$. We define the \emph{prime toric divisors} of $X$ as
	\begin{equation}
		D_p:=\hat{D}_p\cap X\, ,
	\end{equation}
	but note that these complex surfaces in $X$ are not necessarily toric varieties themselves. A choice of K\"ahler form on $V$ induces a K\"ahler form on $X$,
	\begin{equation}
		J_X:=\sum_{p\in \mathcal{P}^\circ}\mathtt{t}_p [D_p]\, ,
	\end{equation}
	so the parameters $\mathtt{t}$ can be viewed as K\"ahler moduli of $X$.\footnote{Note that strictly speaking the prime toric divisors of $X$ are the irreducible components of the $D_p$ defined above. For non-favorably embedded Calabi-Yau hypersurfaces $X$ these still generate $H^2(X,\mathbb{Z})$.} In contrast, variations of the $\psi_q$ control the complex structure of $X$.
	
	\section{Orientifolding Kreuzer-Skarke}\label{sec:orientifolding_KS}
	
	In this section we will describe how to systematically and efficiently enumerate inequivalent holomorphic orientifold involutions of Calabi-Yau threefold hypersurfaces in toric fourfolds constructed from FRSTs of the $4d$ reflexive polytopes in the Kreuzer-Skarke database. We will show how to obtain all such involutions that are inherited from holomorphic involutions of (singular limits) of the toric ambient fourfolds. We will explain how to compute their fixed point sets --- which give rise to the orientifold planes --- and compute their relevant topological data. 
	
	We begin with \S\ref{sec:automorphismsV} by recalling the construction of the automorphism group of a toric fourfold of \cite{Cox:1993fz}. We then go on to classify its inequivalent $\mathbb{Z}_2$ conjugacy classes in \S\ref{sec:Z2conjugacyclassesV}, and describe in \S\ref{sec:invariant_toric_fans} how to obtain $\mathbb{Z}_2$ invariant toric fans from symmetric FRSTs or subdivisions of reflexive polytopes. In \S\ref{sec:fixed_points} we construct the fixed point sets of $\mathbb{Z}_2$ involutions. In \S\ref{sec:Z2automorphismsCY} we discuss how $\mathbb{Z}_2$ involutions of toric ambient varieties induce orientifold involutions of their appropriately tuned Calabi-Yau hypersurfaces, and in \S\ref{sec:smoothness} we detail the relevant smoothness conditions on the generic $\mathbb{Z}_2$ invariant hypersurfaces. In \S\ref{sec:hodgenumbers} we compute the orientifold weighted Hodge numbers. Finally, in \S\ref{sec:orientifolds_at_LCS} we explain how to select involutions that lead to Calabi-Yau hypersurfaces with orientifold invariant large complex structure limits.
	
	\subsection{The automorphism group $Aut(V,\mathbb{C})$}\label{sec:automorphismsV}
	
	The automorphism group $Aut(V,\mathbb{C})$ of a toric variety $V$ has been constructed in \cite{Cox:1993fz}. To specify its elements it is useful to work in homogeneous coordinates  $[x_1:...:x_n]$, i.e. make use of the isomorphism of eq. \eqref{eq:toric_isomorphism}. The component of $Aut(V,\mathbb{C})$ connected to the identity --- which we will denote by $Aut^0(V,\mathbb{C})$ --- is parameterized by maps of the form
	\begin{equation}\label{eq:Aut0}
	x_p\mapsto  \lambda^{p0} x_p+\sum_{q\in (\Theta_p\backslash\del \Theta_p)\cap M}\lambda^{pq} {s_q}^p \, ,\quad \text{with}\quad 
	{s_q}^p:=\prod_{p'\in \mathcal{P}^\circ\backslash\{p\}}x_{p'}^{\langle p',q \rangle}\, ,
	\end{equation}
	modulo the toric scaling relations $G$. 
	
	Here, for vertices $p$ of $\Delta^\circ$ we denote by $\Theta_p$ the dual facet of $\Delta$ and $\Theta_p=\emptyset$ for non-vertex points $p\in \Delta^\circ$. The ${s_q}^p$ are the monomials generating $H^0(V,\mathcal{O}(D_p))$ other than $x_p$, so it is immediately obvious that the above transformations make sense. The subgroup of $Aut^0(V,\mathbb{C})$ given by setting $\lambda^{pq}=0$, and $\lambda^{p0}\neq 0$, is the action of the algebraic torus $T\subset V$, $(T,V)\rightarrow V$.  We denote by $G_{\text{root}}(q)$ --- for $q$ interior to a facet of $\Delta$ --- the one-dimensional subgroups of $Aut^0(V,\mathbb{C})$ generated by maps \eqref{eq:Aut0} with $\lambda^{p0}=1$ and a single non-trivial off-diagonal parameter $\lambda^{pq}$ associated with the point $q\in \Delta$, and call these maps \emph{root automorphisms}.

	The components of $Aut(V,\mathbb{C})$ not connected to the identity are generated composing $Aut^0(V,\mathbb{C})$ with permutations of homogeneous coordinates 
	\begin{equation}\label{eq:AutPerm}
		x_p\mapsto x_{L(p)}\, ,
	\end{equation}
	that are induced by lattice automorphisms $L\in Aut(N)$ that map the triangulated polytope $(\Delta^\circ,\mathcal{T})$ to itself, i.e.
	\begin{equation}\label{eq:def_polytope_automorphism}
		L\in Aut((\Delta^\circ,\mathcal{T}),\mathbb{Z}):=\{L\in Aut(N): \, L(\Delta^\circ)=\Delta^\circ \quad \text{and}\quad L (\mathcal{T})=\mathcal{T}\}\, .
	\end{equation}
	The fact that such maps make sense is perhaps not obvious:  
	the homogeneous coordinates $x_p$ and  $x_{L(p)}$ are not necessarily sections of the same line bundle. Let us briefly explain why this is not a problem. Indeed, an automorphism of $\mathbb{C}^{n+4}$ that permutes homogeneous coordinates in some generic way does not make sense on $\mathbb{C}^{n+4}\backslash Z$ because the exclusion set $Z$ may be in the image of points in $\mathbb{C}^{n+4}$. However, by choosing the automorphism on $\mathbb{C}^{n+4}$ in such a way that each cone $\sigma\in \Sigma_{\mathcal{T}}$ gets mapped to another cone $\sigma'\in \Sigma_{\mathcal{T}}$ one ensures that the automorphism does make sense on $\mathbb{C}^{n+4}\backslash Z$. Yet, it still may not make sense on $V\simeq (\mathbb{C}^{n+4}\backslash Z)/G$ if different representatives in the same $G$-orbit on $\mathbb{C}^{n+4}\backslash Z$ get mapped to points in distinct $G$-orbits. If, however, the permutation of homogeneous coordinates is inherited from a linear symmetry acting on the points in $\Delta^\circ$, as in \eqref{eq:def_polytope_automorphism}, then for every toric scaling relation $\eta\in G$ acting on the homogeneous coordinates there is an $L$-image $\eta'$ acting exactly like $g$ but on the permuted coordinates. In this case, the image of a $G$-orbit in $\mathbb{C}^{n+4}\backslash Z$ is also a $G$-orbit. Thus, the permutations of homogeneous coordinates inherited from linear symmetries of $(\Delta^\circ,\mathcal{T})$ do make sense as automorphisms of $V$.
	
\begin{figure}
		\centering
		\includegraphics[keepaspectratio,width=10cm,clip,trim=6cm 6cm 9cm 7cm]{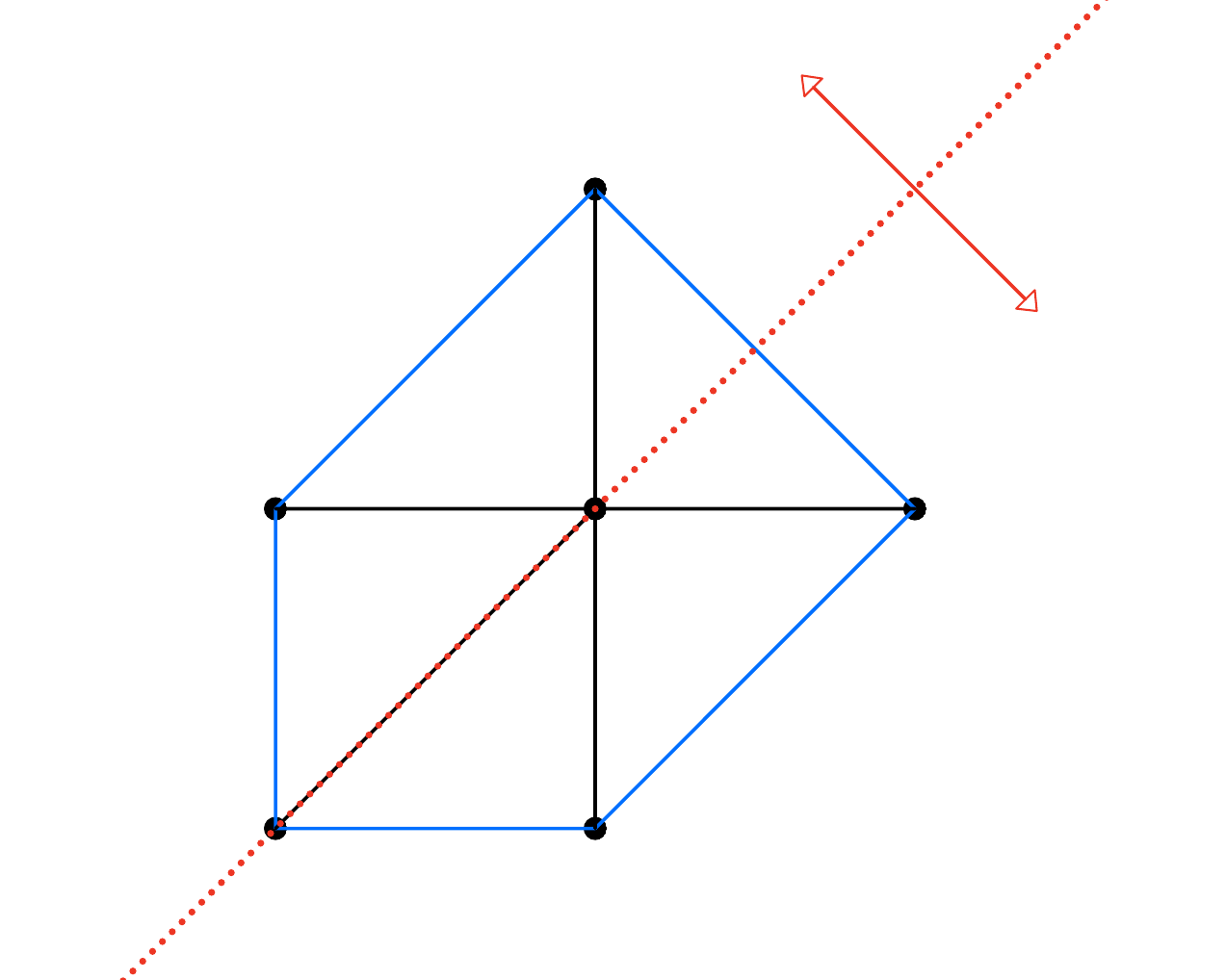}
		\caption{Toric fan of $dP_2$. In red: lattice automorphism defined via reflection across the diagonal.}
		\label{fig:dP2_fan}
	\end{figure}

	To exemplify this let us consider the complex surface $dP_2$, which is $\mathbb{P}^2$ blown up at two generic points. Its toric fan --- depicted in Figure \ref{fig:dP2_fan} 
	is obtained from the polytope $\Delta^\circ$ with points equal to the columns of
	\begin{equation}
		\begin{pmatrix}
			1 & 0 & -1 & -1& 0\\
			0 & 1 & -1 & 0 & -1
		\end{pmatrix}\, ,
	\end{equation}
	associated with five homogeneous toric coordinates $[x_1:\ldots: x_5]$.  The lattice automorphism defined by $L=\begin{pmatrix}
		0 & 1\\
		1 & 0
	\end{pmatrix}$ --- also depicted in Figure \ref{fig:dP2_fan} --- defines an involution
	\begin{equation}\label{eq:dP2_map}
		x_1\leftrightarrow x_2\, ,\quad x_3\mapsto x_3\, ,\quad x_4\leftrightarrow x_5\, .
	\end{equation}
	In order to confirm that this map makes sense on $dP_2$ let us construct a GLSM charge matrix $Q$ collecting the linear relations among the points in $\Delta^\circ$,
	\begin{equation}
		Q=\begin{pmatrix}
			1 &1 & 1 & 0 & 0\\
			0 & 1 & 1 & -1 & 0\\
			1 & 0 & 1 & 0 & -1
		\end{pmatrix}\, ,
	\end{equation}
	whose rows define the weights of the group of toric scaling relations $G_{\text{toric}}\simeq (\mathbb{C}^*)^3$. The map \eqref{eq:dP2_map} acts as a permutation of columns of $Q$, and can be undone by a permutation of the last two rows. Thus, indeed, every pair of homogeneous coordinates $\vec{x}=(x_1,\ldots,x_5)$ and $\vec{x}'=(x_1',\ldots,x_5)$ related by a toric scaling relation $(\lambda_1,\lambda_2,\lambda_3)\in (\mathbb{C}^*)^3$ gets mapped to a pair of homogeneous coordinates related by an image toric scaling relation $(\lambda_1,\lambda_3,\lambda_2)\in (\mathbb{C}^*)^3$.

	We now return to the general discussion. By a slight abuse of notation, and in order to avoid (even more) cluttered notation, we will denote the discrete subgroup of $Aut(V,\mathbb{C})$ induced by elements $L\in Aut((\Delta^\circ,\mathcal{T}),\mathbb{Z})$ also by $Aut((\Delta^\circ,\mathcal{T}),\mathbb{Z})$. Furthermore, we will denote the dual action on points $q\in M$ also by $L(q):= L^T\cdot q$.
	
	We note that the automorphisms \eqref{eq:Aut0} do not generally commute with those of \eqref{eq:AutPerm}. For instance, algebraic torus elements $[t]\in T\subset Aut^0(V,\mathbb{C})$ satisfy the following commutation relation
	\begin{equation}\label{eq:commutation_relation}
		L\circ \phi_{[t]}\circ L^{-1}=\phi_{[L(t)]}\, ,\quad \forall L\in Aut((\Delta^\circ,\mathcal{T}),\mathbb{Z})\, .
	\end{equation}
	Next, in order to understand the induced action of an automorphism on K\"ahler moduli space, we need to understand the induced action of automorphisms on $H^2(V)$. For this, we note that maps of the form \eqref{eq:Aut0}  act trivially on the toric divisor classes $[\hat{D}_p]$ and therefore the induced action of an automorphism on the cohomology group $H^2(V,\mathbb{Z})$ is completely determined by the choice of an element $L\in Aut((\Delta^\circ,\mathcal{T}),\mathbb{Z})$. In particular, we have
	\begin{equation}
		[\hat{D}_p]\mapsto[\hat{D}_{L(p)}]\, ,\quad \hat{D}_p:=\{x_p=0\}\subset V\, . 
	\end{equation}
	As the $[\hat{D}_p]$ generate $H^2(V,\mathbb{Z})$ the above completely determines the induced action on the cohomology group $H^2(V,\mathbb{Z})$, and we will denote it by
	\begin{equation}
		\Lambda_L:\, H^2(V,\mathbb{Z})\rightarrow H^2(V,\mathbb{Z})\, .
	\end{equation}
	Constructing $\Lambda_L$ is thus straightforward.

	\subsection{Enumerating $\mathbb{Z}_2$ conjugacy classes of $Aut(V,\mathbb{C})$}\label{sec:Z2conjugacyclassesV}

	In order to classify orientifold involutions of Calabi-Yau threefold hypersurfaces in toric fourfolds $V$, we must consider automorphisms that square to the identity, i.e. involutions. We are therefore led to consider the subgroup 
	\begin{equation}
		Aut_{\sqrt{\id}}(V,\mathbb{C}):=\{g\in Aut(V,\mathbb{C}):\, g^2=\id \}\subset Aut(V,\mathbb{C})\, .
	\end{equation}
	If two elements of $Aut_{\sqrt{\id}}(V,\mathbb{C})$ are conjugate to each other via an element of $Aut(V,\mathbb{C})$ the corresponding involutions --- and the the resulting orientifolds --- have identical properties. As a result we are really interested in $\mathbb{Z}_2$ conjugacy classes:
	\begin{align}
		\widetilde{Aut}_{\sqrt{\id}}(V,\mathbb{C})&:=Aut_{\sqrt{\id}}(V,\mathbb{C})/\sim \, ,\nonumber\\ 
		\text{with}\quad g\sim g'\quad &\Leftrightarrow: \quad \exists g''\in Aut(V,\mathbb{C}):\quad g'=g''\circ g \circ {g''}^{-1}\, .
	\end{align}
	It is not difficult to see that any such class can be represented by a composition 
	\begin{equation}
		L\circ \phi_{[t]}\, 
	\end{equation}
	of an algebraic torus element $\phi_{[t]}\in T$,
	and a permutation of pairs of homogeneous coordinates 
	\begin{equation}
		L:\, x_p\leftrightarrow x_{L(p)}\, ,
	\end{equation}
	induced by a $\mathbb{Z}_2$-symmetry of the polytope $\Delta^\circ$, in other words by an element of
	\begin{equation}\label{eq:def_automorphisms_polytope_sq1}
		Aut_{\sqrt{\id}}((\Delta^\circ,\mathcal{T}),\mathbb{Z}):=\{L\in Aut((\Delta^\circ,\mathcal{T}),\mathbb{Z}):\,  L^2=\id\}\, .
	\end{equation}
	The reason is as follows. Any $g\in Aut^0(V,\mathbb{C})$ --- i.e. of the form \eqref{eq:Aut0} --- squares to the identity if and only if all the ${s_q}^p$ with $\lambda^{qp}\neq 0$ are themselves homogeneous coordinates. Then, $g$ is a general linear transformation rotating homogeneous coordinates that are sections of the same line bundle. If such $g$ squares to the identity (modulo overall scaling of homogeneous coordinates), it can be diagonalized with non-zero eigenvalues, and therefore its conjugacy class can be represented by a map $g'\in T\subset Aut_0(V,\mathbb{C})$. 
	
	In the following we will explain how to enumerate a set of tuples $(L,\phi_{[t]})$ that represent the inequivalent $\mathbb{Z}_2$ conjugacy classes of $Aut(V,\mathbb{C})$.
	
	First, any $L\in Aut_{\sqrt{\id}}((\Delta^\circ,\mathcal{T}),\mathbb{Z})$ that permutes pairs of homogeneous coordinates $(x_p,x_{L(p)})$ that take values in the same line bundle take values in $Aut^0(V,\mathbb{C})$ and can therefore be diagonalized into an algebraic torus action. We are thus led to define an equivalence relation on $Aut_{\sqrt{\id}}((\Delta^\circ,\mathcal{T}),\mathbb{Z})$,
	\begin{equation}\label{eq:cohom_gauge_fix}
		L\sim L' \quad \Leftrightarrow \quad [D_L(p)]=[D_{L'(p)}] \quad \forall p\in \mathcal{P}^0\, ,
	\end{equation}
	and consider conjugacy classes in $Aut_{\sqrt{\id}}((\Delta^\circ,\mathcal{T}),\mathbb{Z})/\sim$.
	
	Next, given any involution $L\circ \phi_{[t]}$, conjugation via an algebraic torus action $\phi_{[t']}$ leads to the equivalence relation
	\begin{equation}\label{eq:t_gauge_redundancy}
		L\circ \phi_{[t]}\sim \phi_{[t']}\circ L\circ \phi_{[t]}\circ \phi_{[-t']}=L\circ \phi_{t-P_-^L (2t')}\, ,
	\end{equation}
	where we have used the commutation relation \eqref{eq:commutation_relation}, and defined the (anti-)symmetric projection operator by $P_{\pm}^L:=\frac{1}{2}(\id\pm L)$. The continuous part of this redundancy is gauge fixed by imposing that $t$ lives in the invariant subspace of $N\otimes \mathbb{C}$ under $L$. Then, requiring that $L\circ \phi_{[t]}$ squares to the identity implies that
	\begin{equation}\label{eq:squarestoone}
		\id=(L\circ \phi_{[t]})^2=L\circ \phi_{[t]} \circ L^{-1}\circ \phi_{[t]}=\phi_{[t+L(t)]}=\phi_{[2t]}\, ,
	\end{equation}
	i.e. that $2t\in N$. 
	
	Now, for half integral $t'\in \frac{1}{2}N$ in \eqref{eq:t_gauge_redundancy} we have $P_-^L(2t')=P_+^L(2t')\mod N$, so we may further identify
	\begin{equation}
		t_1\sim t_2 \quad \Leftrightarrow\quad t_1-t_2\in  P_+^L(N)\, .
	\end{equation}
	Thus, the inequivalent classes $[2t]$ are the integer points in the coset
	\begin{equation}\label{eq:torusactionconstraint}
		H_+^L:=\frac{P_+^L(N)}{2P_+^L(N)}\, ,
	\end{equation}
	which is isomorphic to $(\mathbb{Z}_2)^k$ with $k\leq 4$.

	At this point, we are left with a residual gauge equivalence
	\begin{equation}
		L\circ \phi_t\sim L'\circ L\circ L'^{-1}\circ \phi_{L'(t)}\, , \quad \forall L\in Aut(\Delta^\circ,\mathbb{Z})\, .
	\end{equation}
	This is partially gauge fixed by considering a single representative $L$ of each $\mathbb{Z}_2$ conjugacy class $[L]$ of $Aut((\Delta^\circ,\mathcal{T}),\mathbb{Z})$ (further identified via \eqref{eq:cohom_gauge_fix}). This gauge fixing prescription then leaves a final residual equivalence $\phi_{[t]}\sim \phi_{[L'(t)]}$ for all $L'$ that commute with $L$. Again this is dealt with trivially by choosing a single representative per class.
	
	At this point we have in hand a minimal set of tuples $(t,L)$ that represent all distinct $\mathbb{Z}_2$ conjugacy classes of $Aut(V,\mathbb{C})$.
	
	\subsection{Invariant toric fans}\label{sec:invariant_toric_fans}
	In the preceding discussion we have considered a reflexive polytope $\Delta^\circ$ together with an FRST $\mathcal{T}$ and have defined the relevant automorphism group $Aut_{\sqrt{\id}}((\Delta^\circ,\mathcal{T}),\mathbb{Z})$ with respect to it, cf. \eqref{eq:def_polytope_automorphism} and \eqref{eq:def_automorphisms_polytope_sq1}. In practice, however, it is much more useful to first consider the full automorphism group $Aut(\Delta^\circ,\mathbb{Z})$ of the reflexive polytope $\Delta^\circ$, without restricting to those respecting any given triangulation $\mathcal{T}$, and ask whether a suitable invariant triangulation exists for the (gauge fixed) set of elements $L\in Aut(\Delta^\circ,\mathbb{Z})$.
	
	In order to do this systematically, we first recall a few well known facts about triangulations of polytopes  (see e.g. \cite{GKZ1,GKZ2,GKZ_book,Triangulations_book}).\footnote{We thank Andres Rios-Tascon for explaining these to us.} An FRST $\mathcal{T}$ of $\Delta^\circ$ is obtained as follows. First, the set of \emph{regular} triangulations is obtained by assigning \emph{heights} $\{h_p\}_{p\in \Delta^\circ\cap N}\in \mathbb{R}^{h^{1,1}(V)+5}$, and lifting all points into five dimensions as
	\begin{equation}
		p\mapsto (p,h_p)\, ,\quad \forall p\in \Delta^\circ\cap N\, .
	\end{equation}
	One then defines the simplices of $\mathcal{T}$ via projection of the lower faces of the $5d$ polytope $\Delta^\circ_5$ defined as the convex hull of the points $(p,h_p)$. Demanding that no point $(p,h_p)$ is interior to  $\Delta^\circ_5$ leads to linear inequalities on the heights, and the resulting triangulations are \emph{fine} (i.e. include all points). For sufficiently low height of the origin $h_0$, the triangulation becomes \emph{star}, i.e. the origin is a vertex of each simplex of $\mathcal{T}$.
	
\begin{figure}
		\centering
		\includegraphics[keepaspectratio,width=10cm,clip,trim=2cm 10cm 2cm 2cm]{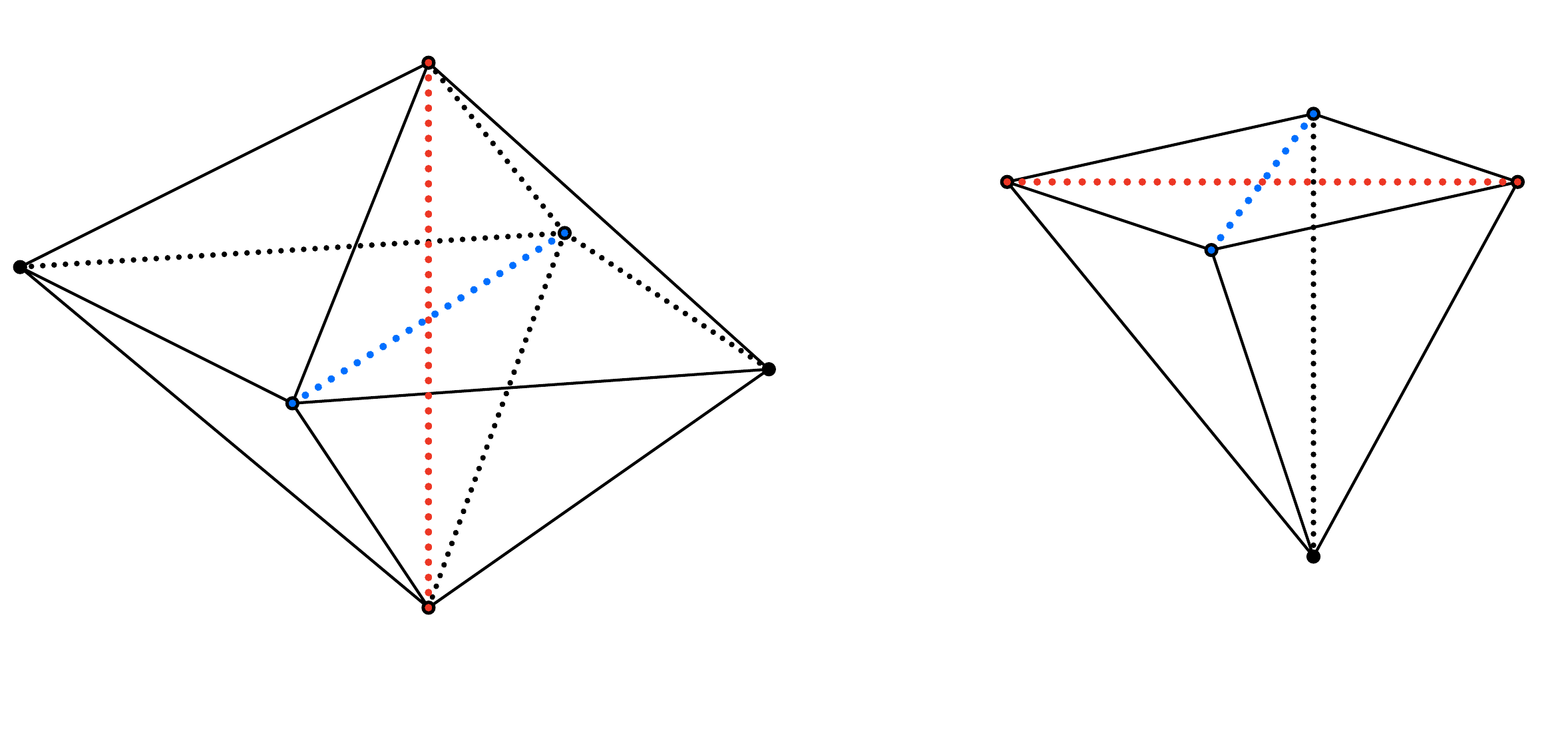}
		\caption{Illustration of bi-stellar flips: as examples we depict two three-faces of a $4d$ reflexive polytope, and their distinct triangulations (blue vs red dotted lines). Left: the bi-stellar flip leaves the induced triangulation of two-faces unchanged, so shrinking the corresponding curve in the toric fourfold leaves the generic Calabi-Yau hypersurface smooth. Right: the bi-stellar flip changes the induced triangulation of the upper two-face, and therefore the generic Calabi-Yau threefold undergoes a flop transition.}
		\label{fig:bistellar_flips}
	\end{figure}

	The set of heights leading to the same FRST of $\Delta^\circ$ forms a convex cone $\hat{\Sigma}_{\mathcal{T}}(\Delta^\circ)$, called the \emph{secondary cone} of the triangulation $\mathcal{T}$, and is related to the K\"ahler cone $\mathcal{K}_V$ of the resulting projective toric variety $V$ via
	\begin{equation}
		\{h_p\}_{p\in \Delta^\circ\cap N}\mapsto J_V=\sum_{p\in \mathcal{P}^\circ}(h_p-h_0)[\hat{D}_p]\, .
	\end{equation}
	The union of secondary cones $\hat{\Sigma}_{\mathcal{T}}(\Delta^\circ)$ over all regular triangulations $\mathcal{T}$ gives the \emph{secondary fan} $\hat{\Sigma}_{\text{RT}}(\Delta^\circ)$, which is a complete fan. Its sub-fan $\hat{\Sigma}_{\text{FRST}}(\Delta^\circ)$ that is formed by height vectors resulting in FRSTs is subdivided into the secondary cones of all the distinct FRSTs by a set of hyperplanes, corresponding to regular subdivisions that aren't triangulations. Crossing these hyperplanes corresponds to bi-stellar flips --- see  Figure \ref{fig:bistellar_flips} 
	--- between pairs of triangulations, and including them makes $\hat{\Sigma}_{\text{FRST}}(\Delta^\circ)$ a convex cone \cite{Triangulations_book}.
	
	Using this, given any $L\in Aut_{\sqrt{\id}}(\Delta^\circ,\mathbb{Z})$ there is a simple prescription that generates an $L$-invariant fine, regular, and star subdivision of $\Delta^\circ$, from any FRST $\mathcal{T}'$: one considers a set of heights $\{h'_p\}$ in the secondary cone of $\mathcal{T}'$, for example those defining a Delaunay triangulation of $\Delta^\circ$, and defines $L$-symmetrized heights
	\begin{equation}\label{eq:symmetrized_heights}
		h_p:=\frac{1}{2}(h'_p+h'_{L(p)})\, .
	\end{equation}
	Due to convexity of $\hat{\Sigma}_{\text{FRST}}(\Delta^\circ)$ this is guaranteed to generate an $L$-invariant fine, regular, and star triangulation $\mathcal{T}$ \emph{or} subdivision $\mathcal{T}_{\text{sub}}$ of $\Delta^\circ$. In the former case one has succeeded in constructing an $L$-invariant toric fan. We note that generating FRSTs $\mathcal{T}$ from heights is essentially instantaneous in \texttt{CYTools} \cite{Demirtas:2022hqf}.
	
	In the following we will assume the latter case, i.e. that the above prescription has resulted only in a fine, regular, and star subdivision, but not a triangulation. Then, by a small generic perturbation of the heights, 
	\begin{equation}
		h_p\mapsto h_p+\delta h_p
	\end{equation}
	one generates an FRST $\mathcal{T}$ of $\Delta^\circ$ whose secondary cone intersects the $L$-symmetric locus along a face corresponding to one or more bi-stellar flips.
	
	Now, any $L$-symmetric choice of K\"ahler class leads to a singular toric variety $V_s$ that can be thought of as a singular limit of the smooth toric variety associated with $\mathcal{T}$, corresponding to the shrinking of a set of rational curves
	\begin{equation}\label{eq:toric_curves_V}
		\hat{\mathcal{C}}_{p,p',p''}:=\hat{D}_p\cap \hat{D}_{p'}\cap \hat{D}_{p''}\, ,
	\end{equation}
	with $p,p',p''\in \mathcal{P}^\circ$ forming a triangle in $\mathcal{T}$. Importantly, the generic Calabi-Yau hypersurface remains smooth in the limit where the toric curves \eqref{eq:toric_curves_V} degenerate --- i.e. in the $L$-symmetric limit --- if and only if the symmetrized heights \eqref{eq:symmetrized_heights} induce fine and regular triangulations of all the two-faces of $\Delta^\circ$ (as in the first example shown in Figure \ref{fig:bistellar_flips}). Otherwise, the generic $\mathbb{Z}_2$-invariant hypersurface is singular (second example shown in Figure \ref{fig:bistellar_flips}).
	
	The toric fan of the singular toric variety $V_s$ is obtained from the one of $V$ by removing the three-dimensional cones $\sigma_{p,p',p''}\in \Sigma_{\mathcal{T}}$ associated to the degenerate curves $\hat{\mathcal{C}}_{p,p',p''}$, and pairwise gluing together of cones that intersect along the $\sigma_{p,p',p''}$. In what follows we will drop the subscript and assume that the toric ambient variety $V$ is defined by an $L$-invariant fan, constructed as explained above, which is induced by an FRST of $\Delta^\circ$ or a fine, regular and star subdivision that induces fine, and regular triangulations of all the two-faces.
	
	\subsection{Fixed points}\label{sec:fixed_points}
	Next, we turn to the fixed point set $\mathcal{F}_{\tilde{\mathcal{I}}}\subset V$ of an involution $\phi_{[t]}\circ L\in Aut_{\sqrt{\id}}(V,\mathbb{C})$.
	
	Every irreducible component of $\mathcal{F}_{\tilde{\mathcal{I}}}$ will turn out to correspond to a toric (but not necessarily torus invariant) subvariety of $V$. To enumerate these we first define an auxiliary toric variety $V_L$, via constructing a toric fan $\Sigma_L$ as follows. Letting $N^L_{\mathbb{R}}$ be the $L$-invariant sub vector space of $N_{\mathbb{R}}$, we set
	\begin{equation}\label{eq:symmetrized_toric_fan}
		\Sigma_L:=\{\sigma \cap N^L_{\mathbb{R}}\}_{\sigma\in \Sigma}\, .
	\end{equation}
	The one-dimensional cones (edges) of $\Sigma_L$ fall into two classes: first, the $L$-symmetric edges of $\Sigma$, and second the intersection of $N_{\mathbb{R}}^L$ with the two dimensional cones in $\Sigma$ generated by pairs $(p,L(p))$ with $p\neq L(p)$.
	
	An intuitive description of $V_L$ is as a subvariety of $V\simeq (\mathbb{C}^{n+4}\backslash Z)/G$ defined by setting
	\begin{equation}\label{eq:VL_defining_eq}
		x_p=x_{L(p)}\quad \forall p\in \mathcal{P}^\circ\, ,
	\end{equation}
	viewed as a set of linear constraints in $\mathbb{C}^{n+4}\backslash Z$. Identifying points related by the group of toric scaling relations $G$ defines $V_L$. Clearly, every point in $V_L\subset V$ is invariant under $L$.
	
	One can equivalently think of $V_L$ as being parameterized by the subset of homogeneous coordinates $x_p$ which are mapped to themselves by $L$, as well as a further homogeneous coordinate for every pair $(x_p,x_{L(p)})$ --- with $L(p)\neq p$ --- parameterizing the diagonal locus $x_p=x_{L(p)}$. The group of toric scaling relations of the toric variety $V_L$ is the subgroup $G_L\subset G$ that leaves the ratios $x_p/x_{L(p)}$ invariant.\footnote{In physics terms the gauge group of toric scaling relations $G$ is broken spontaneously to its subgroup $G_L$ via \eqref{eq:VL_defining_eq}.} Only pairs $(x_p,x_{L(p)})$ that correspond to two dimensional cones of $\Sigma$ need to be considered, as all others cannot vanish simultaneously, and therefore the corresponding diagonal homogeneous coordinate can be gauge fixed. The exclusion set $Z_L$ inherited in this way from $Z$ then indeed corresponds to the one defined by the toric fan of  \eqref{eq:symmetrized_toric_fan}. 
	
	We note that given an $L$-invariant toric fan of $V$ constructed via symmetrization of generic heights as in \ref{sec:invariant_toric_fans} the toric variety $V_L$ constructed here has a simplicial toric fan, and in particular has singularities of at most orbifold type, associated with non-smooth cones in the toric fan $\Sigma_L$.

	Now, let us make an ansatz for a general irreducible component of the fixed point set $\mathcal{F}_{\tilde{\mathcal{I}}}$. We impose
	\begin{equation}\label{eq:ansatz_fixed_locus}
		x_p = e^{2\pi i \eta_p^-}x_{L(p)}\, , \quad \forall p\neq L(p)\, ,
	\end{equation}
	for some fixed $\eta_p^-$, and set to zero the homogeneous coordinates associated with the edges of a pointwise $L$-invariant cone $\sigma\in \Sigma$. Along \eqref{eq:ansatz_fixed_locus} a $\mathbb{Z}_2$ symmetry $L\circ \phi_{[t]}$ --- gauge fixed in the way discussed in the previous section --- acts as
	\begin{align}
		(x_p,x_{L(p)})&\mapsto (e^{2\pi i \eta_p^-}x_p,e^{-2\pi i \eta_p^-}x_{L(p)})\, ,\quad \forall p\neq L(p)\, ,\nonumber\\
		x_p&\mapsto e^{2\pi i \eta_p^t}x_p\, ,\quad \forall p=L(p)\, .
	\end{align}
	with $t=\sum_p \eta_p^t p$.
	
	Now, for our ansatz \eqref{eq:ansatz_fixed_locus} to actually define a fixed point set, one must be able to undo these transformations with a toric scaling relation, composed with an arbitrary scaling symmetry acting on the coordinates that have been set to zero, i.e.
	\begin{equation}\label{eq:pair_phases}
		t+\sum_{p\in \sigma(1)}\alpha_p p+\sum_{\text{pairs }p\neq L(p)}\eta_p^-(p-L(p))\in N\, ,
	\end{equation}
	for some arbitrary $\alpha_p$ defined modulo $\mathbb{Z}$. Applying the anti-symmetric projection operator $P_-^L=\frac{1}{2}(\id-L)$ we conclude that
	\begin{equation}
		\sum_{\text{pairs }p\neq L(p)}\eta_p^-(p-L(p))\in P_-^L(N)\, .
	\end{equation}
	Different choices of the phases $\eta^-_p$ and $\hat{\eta}^-_p$ in our ansatz \eqref{eq:ansatz_fixed_locus} are equivalent if likewise they can be related by a toric scaling relation. This implies the following equivalence relation
	\begin{equation}\label{eq:equivalencerelationsigns}
		\{\eta_p^-\}\sim \{{\eta'}_p^-\}\quad :\Leftrightarrow  \quad 
			\sum_{\text{pairs}}\eta^-_p(p-L(p))-\sum_{\text{pairs}}\hat{\eta}^-_p(p-L(p))\in (\id-L)N\equiv 2P_-^L(N)\, ,
	\end{equation}
	and therefore, choices of phases $\eta_p^-$ defining inequivalent irreducible components of the fixed point set are in one to one correspondence with points $\nu=\sum_{\text{pairs }p\neq L(p)}\eta_p^-(p-L(p))$ in the coset
	\begin{equation}
		H_-^L:=\frac{P_-^L(N)}{2P_-^L(N)}\, ,
	\end{equation}
	that satisfy the integrality condition \eqref{eq:pair_phases}.
	
	Thus, an irreducible component of the fixed point set is naturally labeled by a point-wise $L$-invariant cone $\sigma$, corresponding to the simultaneous vanishing of a set of $L$-invariant homogeneous coordinates, as well as an element
	\begin{equation}
		\nu\in H_-^L\, ,
	\end{equation}
	and we will denote this (toric) variety as $\mathcal{F}_{\tilde{\mathcal{I}}}(\sigma,\nu)$.
	
	Generally, one enumerates the distinct irreducible fixed point set components by iterating over all point wise $L$-invariant cones $\sigma$, and reducing \eqref{eq:pair_phases} modulo the sub vector space spanned by $\sigma$. Then, one removes the $\mathcal{F}_{\tilde{\mathcal{I}}}(\sigma,\nu)$ that are already contained in a higher dimensional component of $\mathcal{F}_{\tilde{I}}$. This is readily implemented.
	
	But in the special case where $\sigma$ is a smooth cone (this will be the relevant case), the following simplification occurs. Without loss of generality, in \eqref{eq:pair_phases} we may set $\alpha_p=\frac{1}{2}$, and one can omit reduction modulo all $L$-symmetric cones.\footnote{To arrive at this one first shows that $\sum_p \alpha_p p\in \frac{1}{2}N$ by applying the operator $\id+L$ to \eqref{eq:pair_phases}. Then, if one of the coefficients $\alpha_p$ is integral, the lower dimensional cone defined by removing the point $p$ already gives rise to a higher dimensional fixed locus component.} 
	
	In summary, the irreducible components $\mathcal{F}_{\tilde{\mathcal{I}}}(\sigma,\nu)$ of the fixed point set $\mathcal{F}_{\tilde{\mathcal{I}}}$ with smooth normal bundle are given by the tuples
	\begin{equation}\label{eq:fixed_loci_summary}
		(\nu,\sigma)\in H_-^L \times  \Sigma\, :\quad  t+\nu+\frac{1}{2}\sum_{p\in \sigma(1)}p\in N\, , 
	\end{equation}
	with $\sigma$ pointwise invariant under $L$.
	
	We return to our example of Figure \ref{fig:dP2_fan}, an involution of the del-Pezzo surface $dP_2$. We have $t=0$ and
	\begin{equation}
		H_-^L=\left\{(0,0)^T\, ,\,\, \frac{1}{2}(1,-1)^T\right\}\, .
	\end{equation}
	The only pointwise invariant cone of the toric fan of $dP_2$ is the one generated by $p=(-1,-1)^T$. Plugging this into \eqref{eq:fixed_loci_summary} we conclude that the fixed point set in $V\simeq dP_2$ has two irreducible components: the first corresponds to the trivial cone $\sigma=0$, and $\nu=0$, and is given by the hypersurface  
	\begin{equation}
		\left\{x_1x_5-x_2x_4=0\right\}\subset V\, .
	\end{equation}
	The other component is an isolated point, corresponding to the one-dimensional cone $\sigma$ generated by $p_3=(-1,-1)^T$ and $\nu=\frac{1}{2}(-1,-1)^T$. It can be represented by the complete intersection 
	\begin{equation}
		\left\{x_1x_5+x_2x_4=x_3=0\right\}\subset V\, .
	\end{equation}
	We note that the fixed point set in this example has previously been computed in \cite{Blumenhagen:2008zz}, yielding the same result.
	
	In general, we also would like to highlight a simplification of the above discussion that occurs when $L=\id$: then, the irreducible components of the fixed point set, with smooth normal directions, correspond to smooth cones $\sigma\in \Sigma$ that satisfy
	\begin{equation}\label{eq:fixed_loci_simplified_formula}
		t+\frac{1}{2}\sum_{p\in \sigma(1)}p\in N\, .
	\end{equation}
	For example, in the commonly studied involutions that can be represented by a sign flip of a single homogeneous coordinate (as e.g. in \cite{Jefferson:2022ssj})
	\begin{equation}
		x_{p_0}\mapsto -x_{p_0}\, ,
	\end{equation}
	one may set $t=\frac{1}{2}p_0$ and readily finds all possible irreducible components of the fixed point set.

	\newpage
	\subsection{Induced orientifolds of Calabi-Yau hypersurfaces}\label{sec:Z2automorphismsCY}
	So far we have classified suitable $\mathbb{Z}_2$ conjugacy classes of holomorphic involutions of toric fourfolds. But really we are interested in their induced action on their suitably generic Calabi-Yau hypersurfaces. 
	
	For an involution $\tilde{\mathcal{I}}\in Aut(V,\mathbb{C})$ to make sense as an element of $Aut(X,\mathbb{C})$ under restriction to the hypersurface, each point on $X$ must be mapped under $\tilde{\mathcal{I}}$ to a point that also lies on $X$. This forces the complex parameters $\psi_q$ in the hypersurface equation \eqref{eq:CY_hypersurface} to be tuned in a way s.t. the vanishing locus of the anti-canonical polynomial $f$ is mapped to itself. In other words,
	\begin{equation}\label{eq:f_invariance}
	f\mapsto \gamma_f\cdot  f\, ,\quad \gamma_f\in \mathbb{C}^*\, .
	\end{equation}
	Writing a general $\mathbb{Z}_2$ involution as $\phi_{[t]}\circ L$ as in \S\ref{sec:Z2conjugacyclassesV} one finds via straightforward evaluation that
	\begin{equation}
		f=\sum_{q\in \Delta\cap M}\psi_q s_q\mapsto e^{2\pi i\sum_p \eta_p}\sum_{q\in \Delta\cap M}\psi_q e^{ 2\pi i\langle t,q\rangle} s_{L(q)}=e^{2\pi i\sum_p \eta_p}\sum_{q\in \Delta\cap M}\psi_{L(q)} e^{2\pi i \langle t,q\rangle } s_{q}\, ,
	\end{equation}
	where $t=\sum_p \eta_p p$ is some arbitrary decomposition of $t$ in terms of the $p\in \Delta^\circ$.
	
	This satisfies \eqref{eq:f_invariance} if and only if the complex parameters $\psi_q$ satisfy the following constraints
	\begin{equation}\label{eq:orientifold_invariance_f}
		\psi_{L(q)}\overset{!}{=} e^{2\pi i \left(\langle t,q\rangle+\frac{1}{2}\lambda_f\right)}\psi_q\, ,\quad \forall q\in \Delta\, .
	\end{equation}
	for a fixed choice $\lambda_f\in \mathbb{Z}_2$ that becomes part of the data defining an involution of the Calabi-Yau hypersurface.
	
	Writing the holomorphic threeform of $X$ as a Poincar\'e residue this implies that $\Omega\mapsto \pm \Omega$, corresponding to either an O3/O7 orientifold or an O5/O9 orientifold depending on the choice of $\lambda_f$.
	
	\subsection{Smoothness conditions on invariant hypersurfaces}\label{sec:smoothness}
	In this paper we restrict to $\mathbb{Z}_2$ orientifolds of smooth Calabi-Yau hypersurfaces. The Calabi-Yau hypersurfaces that arise in the Kreuzer-Skarke dataset are smooth at generic points in their moduli space of K\"ahler structures and complex structures. But as discussed in \S\ref{sec:Z2automorphismsCY} a holomorphic $\mathbb{Z}_2$ involution inherited from a toric ambient variety $V$ leads to a holomorphic involution of the hypersurface only if the complex parameters defining the hypersurface are tuned, as in \eqref{eq:orientifold_invariance_f}. Smoothness of the resulting non-generic hypersurface is then no longer guaranteed. Furthermore, as discussed in \S\ref{sec:invariant_toric_fans}, an involution with non-trivial induced action on $H^{2}(V,\mathbb{Z})$ is a symmetry of the ambient variety viewed as a K\"ahler manifold if and only if the K\"ahler class $J$ is $\mathbb{Z}_2$ invariant. The $\mathbb{Z}_2$ invariant subspace of $H^{2}(V,\mathbb{Z})$ may or may not intersect the K\"ahler cone.

	Thus, the smoothness condition can be split into two separate components, 
	\begin{enumerate}
		\item the $h^{1,1}_+$-dimensional hyperplane of $\mathbb{Z}_2$ invariant K\"ahler classes intersects the K\"ahler cone $\mathcal{K}_X$, and
		\item the generic $\mathbb{Z}_2$-invariant hypersurface is smooth as a complex variety.
	\end{enumerate}
	We will call a Calabi-Yau hypersurface that meets these respective conditions \emph{K\"ahler smooth}, and \emph{complex structure smooth}.
	
	By virtue of the discussion of \S\ref{sec:invariant_toric_fans} a Calabi-Yau hypersurface is K\"ahler smooth if and only if generic heights in the $L$-invariant locus of $\hat{\Sigma}_{\text{FRST}}(\Delta^\circ)$ induce fine and regular triangulations of all the two-faces of $\Delta^\circ$. Precisely in this case, the ambient variety, equipped with a generic $\mathbb{Z}_2$ invariant K\"ahler class, is smooth in the vicinity of the generic hypersurface. We would like to emphasize that admitting singular $\mathbb{Z}_2$ symmetric toric varieties whose Calabi-Yau hypersurface nevertheless remains smooth is key to finding orientifolds with $L\neq \id$ in the Kreuzer-Skarke dataset: while for small $h^{1,1}\lesssim 10$ one can find smooth $L$-invariant toric ambient varieties (or toric orbifolds) with $L\neq \id$, at large $h^{1,1}$ all examples known to us feature non-simplicial $L$-symmetric toric fans, see for example the example \S\ref{sec:largeh11minus}.
	
	Next, given a K\"ahler smooth Calabi-Yau hypersurface $X$, one needs to determine whether its generic $\mathbb{Z}_2$ invariant complex structure still amounts to a smooth hypersurface, with smooth fixed point set under the involution.
	
	First, while a general Calabi-Yau hypersurface in the Kreuzer-Skarke set is smooth, upon imposing constraints on the polynomial coefficients as in \eqref{eq:f_invariance} there may exist solutions to the otherwise overconstrained critical point equations $df=f=0$. One simple way these can arise is if the coefficient of a monomial associated with a vertex of $\Delta$ vanishes for orientifold invariant complex structure \eqref{eq:f_invariance}, and the dual facet of $\Delta^\circ$ has at least one interior point $p$. Then, $f$ factorizes
	\begin{equation}
		f=x_p\cdot f'\, ,
	\end{equation}
	and the hypersurface is singular along the surface $x_p=f'=0$. Thus, we impose the following necessary condition
	\begin{equation}
		 2\langle t,q \rangle+\lambda_f\overset{!}{=}0\mod 2\quad \forall \text{vertices $q\in \Delta$ such that }L(q)=q\, .
	\end{equation}
	We impose the same constraint for all vertices dual to facets of $\Delta^\circ$ that intersect non-simplicial and/or non-smooth cones of the $\mathbb{Z}_2$ symmetric toric fan $\Sigma$, in order for the Calabi-Yau hypersurface to avoid the corresponding singularities.

	Next, we inspect the generic orientifold covariant polynomial $f$ along an irreducible component of the fixed point set $\mathcal{F}_{\tilde{\mathcal{I}}}(\sigma,\nu)$.
	
	We have that
	\begin{equation}
		f|_{\mathcal{F}_{\tilde{\mathcal{I}}}(\sigma,\nu)}\equiv 0\quad \text{if}\quad \text{dim}(\sigma)+\lambda_f=1\mod 2\, ,
	\end{equation}
	while otherwise $f|_{\mathcal{F}_{\tilde{\mathcal{I}}}(\sigma,\nu)}$ is a generic global section of a line bundle on $\mathcal{F}_{\tilde{\mathcal{I}}}(\sigma,\nu)$. In particular, we see that for fixed choice of $\lambda_f$ the complex dimensions of the induced fixed loci in the Calabi-Yau hypersurface
	\begin{equation}\label{eq:fixed_points_X}
		\mathcal{F}_{\mathcal{I}}(\sigma,\nu):=\mathcal{F}_{\tilde{\mathcal{I}}}(\sigma,\nu)\cap X
	\end{equation}
	are either all even or all odd, as they must for the involution to have a consistent induced action on the holomorphic threeform of $X$. As the $\mathcal{F}_{\mathcal{I}}(\sigma,\nu)$ are either toric varieties or effective divisors therein it is in general straightforward to compute their relevant data.
	
	If $f|_{\mathcal{F}_{\tilde{\mathcal{I}}}(\sigma,\nu)}\equiv 0$ for co-dimension one $\mathcal{F}_{\tilde{\mathcal{I}}}(\sigma,\nu)\subset V$, the polynomial $f$ again factorizes, so we further demand that   
	\begin{equation}
		\text{dim}(\sigma)+\lambda_f\overset{!}{=}0\mod 2\quad \forall \, \mathcal{F}_{\tilde{\mathcal{I}}}(\sigma,\nu) \text{ of complex dimension three}\, .
	\end{equation}
	Next, let $\mathcal{F}_{\tilde{\mathcal{I}}}(\sigma,\nu)$ be of complex dimension two. If $\text{dim}(\sigma)+\lambda_f=0\mod 2$ then  $\mathcal{F}_{\mathcal{I}}(\sigma,\nu)$ is a complex curve in $X$ (the location of an O5 plane). If on the other hand $\text{dim}(\sigma)+\lambda_f=1\mod 2$ we have that $\mathcal{F}_{\mathcal{I}}(\sigma,\nu)=\mathcal{F}_{\tilde{\mathcal{I}}}(\sigma,\nu)$ is a complex surface $\mathcal{S}\subset X\subset V$.  Such fixed surfaces in the hypersurface $X$ generically contain isolated nodal points of $X$ \cite{Carta:2020ohw}: the differential $df|_{\mathcal{S}}$ is naturally a section of $\mathcal{O}(K_V^{-1})|_{\mathcal{S}}\otimes \mathcal{N}^* \mathcal{S}$ where $K_V^{-1}$ is the anti-canonical line bundle on $V$. Thus we find singular points $df=f=0$ along a number of isolated points on $\mathcal{S}$ counted by
	\begin{equation}
		n^{\mathcal{S}}_{df=0}=\int_{\mathcal{S}}  c_2(\mathcal{O}(K_V^{-1})|_{\mathcal{S}}\otimes \mathcal{N}^* \mathcal{S})\, .
	\end{equation}
	To avoid such singularities we additionally impose that
	\begin{equation}
		n^{\mathcal{S}}_{df=0}\overset{!}{=}0 \quad \forall \, \mathcal{F}_L(\sigma,\nu) \text{ of complex dimension two with }\text{dim}(\sigma)+\lambda_f=1\mod 2\, .
	\end{equation}
	There are no analogous smoothness constraints associated with $\mathcal{F}_{\tilde{\mathcal{I}}}(\sigma,\nu)$ of complex dimension one and zero.
	
	Finally, we impose that the fixed point set $\mathcal{F}_{\mathcal{I}}$ is smooth itself. To this end we impose that the toric varieties $\mathcal{F}_{\tilde{\mathcal{I}}}(\sigma,\nu)$ along which the defining polynomial $f$ vanishes identically, are smooth. For all other $\mathcal{F}_{\tilde{\mathcal{I}}}(\sigma,\nu)$ we impose instead that the generic section of the relevant line bundle is nef, and that no orbifold singularities of $\mathcal{F}_{\tilde{\mathcal{I}}}(\sigma,\nu)$ intersect the hypersurface.
	
	In summary, we have imposed a number of conditions for the general $\mathbb{Z}_2$ invariant Calabi-Yau hypersurface to be smooth. We are aware of a proof that these conditions are in general sufficient, but in practice we have never found singular generic orientifold invariant hypersurfaces that pass the above conditions.
	
	\subsection{Orientifold weighted Hodge numbers}\label{sec:hodgenumbers}
	In this section we compute the orientifold weighted Hodge numbers $(h^{1,1}_\pm(X,\mathcal{I}),h^{2,1}_\pm(X,\mathcal{I}))$. We note that these (as well as the orientifold weighted Hodge numbers of divisors in $X$) have been computed under the simplifying assumption that $L=\id$ and $2t\in \mathcal{P}^\circ$ in \cite{Jefferson:2022ssj}. Here, we will compute the $h^{p,q}_\pm (X,\mathcal{I})$ for general involutions $\tilde{\mathcal{I}}=\phi_{[t]}\circ L$, but leave the computation of the weighted Hodge numbers of divisors in $X$ for future work.
	
	For favorably embedded Calabi-Yau hypersurfaces $X$ in toric fourfolds $V$, in the sense that the natural inclusion $H_2(X,\mathbb{Z})\hookrightarrow H_2(V,\mathbb{Z})$ is an isomorphism, it is straightforward to compute the orientifold weighted Hodge numbers $h^{1,1}_{\pm}(X,\mathcal{I})$ and $h^{2,1}_{\pm}(X,\mathcal{I})$. The dimensions of the even/odd eigenspaces of the induced action $\Lambda_L:\, H^2(V)\rightarrow H^2(V)$ compute the Hodge numbers $h^{1,1}_{\pm}$. Using $h^{1,1}_-$ one computes $h^{2,1}_-$ using the Lefschetz fixed point theorem
	\begin{equation}\label{eq:Lefschetz}
		h^{2,1}_-=h^{1,1}_-+\frac{\chi(\mathcal{F}_{\mathcal{I}})-\chi(X)}{4}-1\, ,
	\end{equation}
	where again $\chi$ denotes the topological Euler characteristic. The latter is computed straightforwardly using our discussion of \S\ref{sec:fixed_points}, specifically \eqref{eq:fixed_points_X}.
	
	For non-favorable hypersurface, following Batyrev \cite{Batyrev:1993oya}, one constructs a set of $h^{1,1}(X)+4$ generators of $H^2(X,\mathbb{Z})$ from the irreducible components of the $D_p:=\hat{D}_p\cap X$: let $p$ be in the interior of a two-face $\Theta_2^\circ$ of $\Delta^\circ$, and let $\Theta_1$ be the dual one face, bounded by two vertices $q_1,q_2\in \Delta$. Then, the generic anti-canonical polynomial $f$ satisfies  
	\begin{equation}
		f|_{x_p=0}=\prod_p x_p^{\langle p,q_1\rangle +1}\times \sum_{k=0}^{g+1}\psi_{k} y^k\, ,
	\end{equation}
	where $\psi_k:=\psi_{q_1+k\cdot \Delta q}$, and $y:=\prod_p x_p^{\langle p , \Delta q \rangle}$, in terms of $\Delta q:=\frac{q_2-q_1}{g+1}\in M$. Here, $g$ is the genus of the two-face $\Theta^\circ_2$, defined as the number of points interior to the dual one-face $\Theta_1$. For generic $\psi_k$ there are $g+1$ distinct solution branches for $y$ and thus one obtains $g+1$ irreducible disjoint components
	\begin{equation}
		D_p=\dot{\bigcup}_{k=1}^{g+1}D_p^{(k)}\, .
	\end{equation}
	For $p\neq L(p)$ all pairs get mapped into each other: 
	\begin{equation}
		D_p^{(i)}\leftrightarrow D_{L(p)}^{(i)}\, ,\quad i=1,\ldots,g+1\, .
	\end{equation} 
	In contrast, if $p=L(p)$ the dual one-face $\Theta_1$ is mapped to itself, either point-wise, or else the bounding vertices are interchanged by the involution $L$. 
	
	In the former case --- i.e. $L(q_{1})=q_1$ and $L(q_2)=q_2$ --- the involution $\tilde{\mathcal{I}}=\phi_{[t]}\circ L$ acts as
	\begin{equation}
		y\mapsto e^{2\pi i \langle t, \Delta q \rangle}y=\pm y\, ,
	\end{equation}
	and so the components $D_p^{(i)}$ get mapped into each other in pairs for $\langle t, \Delta q \rangle\notin \mathbb{Z}$ while they are mapped to themselves if $\langle t, \Delta q \rangle\in \mathbb{Z}$.
	
	In the latter case --- i.e. $L(q_1)=q_2$ --- the involution instead acts according to
	\begin{equation}
		y\mapsto y^{-1}\, .
	\end{equation}
	Setting, for definiteness, to zero the $\psi_k$ associated with points interior to the one-face $\Theta_1$ the $D_p^{(k)}$ correspond to the solutions to
	\begin{equation}
		y^{g+1}=e^{2\pi i \left(\frac{1}{2}\lambda_f+\frac{1}{2}+\langle t, q_1 \rangle\right)}\, .
	\end{equation}
	If $g+1$ is odd the $D^{(k)}$ organize into $g/2$ pairs that are interchanged by the involution, and a single invariant class. If $g+1$ is even, and also $\frac{1}{2}\lambda_f+\frac{1}{2}+\langle t, q_1 \rangle\in \mathbb{Z}$, then two classes are invariant, corresponding to the solutions $y=\pm 1$, while all others come in pairs. If, finally, $g+1$ is even, but $\frac{1}{2}\lambda_f+\frac{1}{2}+\langle t, q_1 \rangle\notin \mathbb{Z}$ then all the $D^{(i)}_p$ come in non-trivial pairs.
	
	The $D_p$ with $p$ not interior to any two-face together with the irreducible components $D^{(i)}_p$ associated with points $p$ interior to two-faces give the desired $h^{1,1}(X)+4$ generators of $H^2(X,\mathbb{Z})$, and therefore our above discussion enables computing the induced action $\Lambda_L$ on $H^2(X)$ also for non-favorably embedded Calabi-Yau hypersurfaces.\footnote{Starting from a GLSM charge matrix $Q$ whose columns are the classes $[D_p]$ expressed in a basis of $H^{2}(V,\mathbb{Z})$ one can construct an expanded $h^{1,1}(X)$ by $h^{1,1}(X)+4$ matrix $\hat{Q}$ whose columns are the above generators of $H^2(X,\mathbb{Z})$ expressed in a basis.} As in the favorable case from this one computes the orientifold-odd Hodge number $h^{1,1}_-(X,\mathcal{I})$, and uses the index theorem to compute $h^{2,1}_-(X,\mathcal{I})$.
	
	\subsection{Orientifolds at Large Complex Structure}\label{sec:orientifolds_at_LCS}
	
	Calabi-Yau hypersurfaces in toric varieties have maximally unipotent singularities in their complex structure moduli space called \emph{large complex structure} (LCS) points, which are the mirror duals of large volume limits \cite{Candelas:1990rm,Candelas:1993dm,Hosono:1993qy,Candelas:1994hw,Hosono:1994ax}. Complex structure moduli space is generically understood in a (weak coupling) expansion around LCS points \cite{Hosono:1993qy,Hosono:1994ax,Demirtas:2023als}, and therefore one would like to understand the conditions under which the orientifold invariant part of moduli space contains such points. We will refer to an orientifold that meets these conditions as an \emph{LCS orientifold}. Holomorphic LCS orientifolds of type IIB string theory are equivalent via mirror symmetry to suitable anti-holomorphic Calabi-Yau orientifolds of type IIA string theory at LCS, see \cite{Grimm:2004ua}.
	
	Thus, somewhat more precisely, we would like to determine the conditions under which orientifold invariant LCS points exist that correspond to large volume limits of a mirror threefold that can be described as a hypersurface in a toric variety arising from an FRST (or subdivision) of the dual polytope $\Delta$ as in Batyrev's construction of mirror symmetry of Calabi-Yau threefold hypersurfaces \cite{Batyrev:1993oya}.
	
	The mirror map relates the complex parameters $\psi_q$ appearing in the defining polynomial $f$ of $X$ to the K\"ahler form $\tilde{J}$ of the mirror threefold $\tilde{X}$ and the flat NS-NS two form $\tilde{B}$ of type IIA string theory as follows. Expanding 
	\begin{equation}
		\tilde{B}+i\tilde{J}=\sum_q u^q [\tilde{D}_q]\, ,
	\end{equation}
	in terms of complex parameters $u^q$ and the prime toric divisors $\tilde{D}_q$ of the mirror threefold (we define the canonical divisor $[\tilde{D}_0]:=-\sum_{q\neq 0}[\tilde{D}_q]$), the complex coefficients $u^q$ are related to the complex parameters $\psi_q$ appearing in the defining polynomial of $X$ \eqref{eq:CY_hypersurface}. In a gauge where $\psi_q=0$ for all $q$ interior to facets of $\Delta$ the mirror map reads \cite{Candelas:1990rm}
	\begin{equation}\label{eq:mirror_map}
		u^q=\frac{\log(\psi_q)}{2\pi i}+\text{holomorphic}\, .
	\end{equation}
	and the holomorphic piece is parametrically sub-leading in an LCS limit of $X$. Neglecting the holomorphic piece, the orientifold constraints \eqref{eq:f_invariance} then imply that
	\begin{equation}\label{eq:symmetrized_heights_dual}
		h^{L(q)}\overset{!}{=}h^q\, ,
	\end{equation}
	where $h^{q}:=\text{Im}(u^q)$ are interpreted as the heights of points in $\Delta$ defining a regular triangulation (or subdivision) of $\Delta$. One can always find solutions to this equation by symmetrizing generic heights as in \S\ref{sec:invariant_toric_fans}, now applied to the mirror polytope $\Delta$. This, as in \cite{Batyrev:1993oya}, defines a ``mirror" toric variety $\tilde{V}$, and  $\tilde{X}$ is its Calabi-Yau hypersurface. 
	
	For involutions with $\lambda_f=0$ and $t=0$ but arbitrary lattice symmetry $L$ the phase of \eqref{eq:f_invariance} merely enforces that the anti-symmetric part of the NS-NS two form $\tilde{B}$ vanishes, so such orientifolds contain LCS points if and only if the symmetrized heights $h^q$ induce fine, regular triangulations of all the two faces of $\Delta$ as in \S\ref{sec:invariant_toric_fans}.
	
	In contrast, however, for general $(t,\lambda_f)$ the constraint \eqref{eq:f_invariance} enforces the vanishing of monomial coefficients
	\begin{equation}\label{eq:vanishing_coefficnets}
		\psi_q=0\quad \forall q:\,\, q=L(q)\quad \text{and} \quad \langle t, q \rangle+\lambda_f=\frac{1}{2}\mod \mathbb{Z}\, .
	\end{equation}
	If the polytope $\Delta$ has no points interior to facets, whenever any coefficient $\psi_q$ vanishes, its corresponding height $h^q$ diverges, so the orientifold even moduli space cannot contain LCS points that are described by FRSTs of $\Delta$. For such polytopes $\Delta$ the only LCS orientifolds are those with $t=0$ and $\lambda_f=0$.
	
	If, more generally, the polytope $\Delta$ does have points interior to facets, our gauge fixing prescription of \S\ref{sec:Z2conjugacyclassesV} is not in general compatible with the gauge fixing prescription in which the mirror maps takes the simple form \eqref{eq:mirror_map} (where all $\psi_q$ corresponding to points interior to facets have been set to zero). In order to transform a general orientifold covariant polynomial $f$ into one that satisfies the gauge fixing prescription used for applying the mirror map, one applies the root automorphisms $G_{\text{root}}(q)$, as defined below \eqref{eq:Aut0}, in order to set to zero the coefficients $\psi_q$ corresponding to points $q$ interior to facets. 
	
	Given an invariant point $q=L(q)$ interior to a facet of $\Delta$, the action of the root automorphism sub-group $G_{\text{root}}(q)\subset Aut^0(V,\mathbb{C})$ leaves the constraints \eqref{eq:f_invariance} intact if and only if it commutes with the involution $\tilde{\mathcal{I}}$, i.e. when $\langle q,t\rangle=0\mod\mathbb{Z}$. On the other hand, for every $q\neq L(q)$ one can define a pair of (anti-)symmetric one-parameter sub-groups of $Aut^0(V,\mathbb{C})$,
	\begin{equation}
		G_{\text{root},\pm}^L(q):=\left\{\Psi_q(\lambda):\, x_p\mapsto x_p+\lambda {s_p}^q\, ,\quad x_{L(p)}\mapsto x_{L(p)}\pm \lambda {s_{L(p)}}^{L(q)}\right\}\, ,
	\end{equation}
	one of which commutes with $\tilde{\mathcal{I}}$, and the other does not.
	
	One obtains an LCS-orientifold from $\tilde{\mathcal{I}}=\phi_{[t]}\circ L$  if and only if the number of points $q\in \Delta$ whose coefficients $\psi_q$ are projected out via \eqref{eq:vanishing_coefficnets} is equal to the sum of the number of root automorphism groups $G_{\text{root}}(q=L(q))$ and $G_{\text{root},\pm}(q\neq L(q))$ that do \emph{not} commute with the involution $\tilde{\mathcal{I}}=\phi_{[t]}\circ L$. 
	
	The latter is computed by
	\begin{align}\label{eq:dimGrootmodHrootI}
		&\#\left(\text{invariant points $q=L(q)$ interior to facets of $\Delta$ s.t. $\langle q,t \rangle=\frac{1}{2}\mod \mathbb{Z}$}\right)\nonumber\\
		&+\#\left(\text{pairs of points $q_1\neq q_2=L(q_1)$ interior to facets of $\Delta$}\right)\, .
	\end{align}
	Therefore, in general, the orientifold invariant sublocus in moduli space intersects (at least the closure of) a large complex structure cone in moduli space if and only if \eqref{eq:dimGrootmodHrootI} is  equal to the number of coefficients $\psi_q$ that are projected out via \eqref{eq:vanishing_coefficnets}.
	
	If furthermore the generic symmetrized heights $h^q$ in \eqref{eq:symmetrized_heights_dual} induce fine, and regular triangulations of all the two faces of $\Delta$  then the orientifold invariant locus in moduli space intersects the interior of an LCS cone. Otherwise, the orientifold invariant slice through moduli space features shrunken mirror curves $\tilde{\mathcal{C}}$ (these are conifolds in $\tilde{X}$), a regime in complex structure moduli space corresponding to a face of an LCS cone of $X$ --- termed coni-LCS --- that has been explored systematically in \cite{Demirtas:2020ffz,Alvarez-Garcia:2020pxd}. Even in this latter case the generic orientifold is smooth if and only if $\int_{\tilde{\mathcal{C}}} \tilde{B}=\frac{1}{2}\mod \mathbb{Z}$ for all shrunken mirror curves $\tilde{\mathcal{C}}$. Otherwise the type IIB orientifold features degenerate three-spheres inducing conifold singularities in $X$. This condition on the $\tilde{B}$-field is readily computed: the vanishing classes $[\tilde{\mathcal{C}}]$ are computed by the edges of an adjacent fine, regular two-face triangulation that are removed upon choosing symmetric heights $h^q$, and these classes are anti-symmetric under the induced action of $L$. The anti-symmetric part of the NS-NS two form $\tilde{B}$ is uniquely determined, modulo integers, by \eqref{eq:f_invariance}, yielding the phases $e^{2\pi i \int_{\tilde{\mathcal{C}}} \tilde{B}}$ unambiguously.
	
	Finally, we would like to point out a useful simplification of the above discussion that applies to the special case $L=\id$, i.e. when the involution $\mathcal{I}$ of $X$ is inherited from an algebraic torus element $\phi_{[t]}\in Aut(V,\mathbb{C})$.\footnote{We would like to thank Manki Kim for collaboration on this point, and allowing us to produce our joint results about ``trilayer" polytopes here.} First, the half-integral $t\in \frac{1}{2}N$ defines a $\mathbb{Z}_2$-grading of the dual polytope $\Delta$,
	\begin{equation}
		\psi_{[t]}:\, q\mapsto e^{2\pi i \langle q,t\rangle}=\pm 1\, ,\quad q\in \Delta\cap M\, .
	\end{equation}
	The involution $\mathcal{I}$ defines an LCS orientifold if and only if 
	\begin{equation}\label{eq:LCS_condition_L=0}
		\#(q:\, \psi_{[t]}(q)=e^{\pi i (\lambda_f-1)})=\#(q\text{ interior to facets of }\Delta: \psi_{[t]}(q)=-1)\, .
	\end{equation}
	This condition is satisfied if the following two conditions hold: 
	\begin{enumerate}
		\item $\lambda_f=1$.
		\item $\Delta$ is the convex hull of a facet $\Theta\subset \Delta$ and a vertex $q_0\notin \Theta$, such that \\$\psi_{[t]}(\Theta)~=~\psi_{[t]}(q_0)~=~-1$.
	\end{enumerate}
	Then, without loss of generality we can assume that $t=\frac{1}{2}p_0$ where $p_0$ is the vertex of $\Delta^\circ$ that is dual to the facet $\Theta\subset \Delta$, and we may represent  
	\begin{equation}
		\tilde{\mathcal{I}}:\, x_{p_0}\mapsto -x_{p_0}\, .
	\end{equation}
	Under this assumption the polytope $\Delta$ has three layers distinguished by $\langle \bullet,p_0\rangle\in \{-1,0,1\}$ and the middle layer with $\langle \bullet,p_0\rangle=0$ corresponds to the monomials that are projected out. The points interior to the facet $\Theta$ are precisely the middle layer points translated by $-q_0$ and therefore \eqref{eq:LCS_condition_L=0} holds. We refer to a polytope $\Delta$ that satisfies these conditions as \emph{trilayer}.\footnote{It is not hard to see that the dual of a trilayer polytope is also trilayer.} 
	
	Orientifolds of this type inherit the entire moduli space of inequivalent embeddings $X\subset V$, but if the embedding $X\hookrightarrow V$ is non-favorable, cf. \S\ref{sec:hodgenumbers}, one may still find $h^{1,1}_+(X,\mathcal{I})<h^{1,1}(X)$ and/or $h^{2,1}_-(X,\mathcal{I})<h^{2,1}(X)$. 
	
	Finally, we note that in the limit where all toric divisors associated with points in the middle layer are blown down, the orientifold $X/\mathcal{I}$ becomes a \emph{toric} threefold with toric fan inherited from the facet of $\Delta^\circ$ with $-p_0$ viewed as the origin. This makes it easy to relate these orientifolds to their F-theory uplifts as in \cite{Collinucci:2008zs,Collinucci:2009uh,Jefferson:2022ssj}, which are then given by genus one fibrations over toric threefolds (or blow-ups thereof).

	\section{Examples}\label{sec:examples}
	\subsection{An orientifold with $(h^{1,1}_+,h^{1,1}_-,h^{2,1}_+,h^{2,1}_-)=(491,0,0,11)$}\label{sec:491}
	As our first example, we consider the polytope $\Delta^\circ$ resulting in the largest known Hodge number $h^{1,1}(X)=h^{1,1}(V)=491$. Its vertices $(p_1,\ldots,p_5)$ are the columns of
	\begin{equation}
		\begin{pmatrix}
			  -63&   0&  0&   1&  21\\
			  -56&   0&  1&   0&  28\\
			  -48&   1&  0&   0&  36\\
			  -42&   0&   0&   0&  42
		\end{pmatrix}\, .
	\end{equation}
	We define a toric fourfold $V$ via the Delaunay triangulation $\mathcal{T}$ of $\Delta^\circ$, constructed using $\mathtt{CYTools}$ \cite{Demirtas:2022hqf}.

	This polytope is trilayer with $p_0\equiv p_1$, and we define an involution $\tilde{\mathcal{I}}$ on the toric ambient variety $V$ via $t=\frac{1}{2}p_1$, and $L=\id$. We define an induced involution $\mathcal{I}$ by setting $\lambda_f=1$. The embedding $X\hookrightarrow V$ is favorable, and thus we find 
	\begin{equation}
		h^{1,1}_+(X,\mathcal{I})=h^{1,1}(X)=491\, .
	\end{equation}
	All local deformations in complex structure of $X$ are obtained by perturbing a generic polynomial $f$, and due to the trilayer property we find
	\begin{equation}
		h^{2,1}_-(X,\mathcal{I})=h^{2,1}(X)=11\, .
	\end{equation}
	The fixed point set $\mathcal{F}_{\mathcal{I}}\subset X$ is a union of many disjoint irreducible components, corresponding to O7 planes wrapped on $118$ prime toric divisors $D_p$, namely those that satisfy \eqref{eq:fixed_loci_simplified_formula}, in this case. $p+p_1\in 2N$. Furthermore, there are $193$ O3 planes, corresponding to the three-dimensional cones $\sigma_{pp'p''}\in \Sigma_{\mathcal{T}}$ from triplets of points $(p,p'p'')$ that satisfy $p+p'+p''+p_1\in 2N$.
	
	Via direct computation, one finds
	\begin{equation}
		\chi(\mathcal{F})=1008\, ,
	\end{equation}
	which, using \eqref{eq:Lefschetz}, confirms our prediction of the orientifold weighted Hodge numbers. The D3-tadpole is
	\begin{equation}
		Q_{D3}=\frac{\chi(\mathcal{F})}{4}=252=\frac{1}{2}(h^{1,1}(X)+h^{2,1}(X))+1\, ,
	\end{equation}
	the largest possible value in the Kreuzer-Skarke database.\footnote{An analogous orientifold, with the same D3-tadpole, of the mirror dual Calabi-Yau with $h^{1,1}(X)=11$ has been detailed in \cite{Crino:2022zjk}.} No other orientifolds satisfying our smoothness criteria of \S\ref{sec:smoothness} are inherited from involutions of toric ambient fourfolds.

	\subsection{An orientifold with $(h^{1,1}_+,h^{1,1}_-,h^{2,1}_+,h^{2,1}_-)=(123,120,0,3)$}\label{sec:largeh11minus}
	As our second example, we consider the reflexive polytope $\Delta^\circ$ whose vertices $(p_1,\ldots,p_6)$ are the columns of
	\begin{equation}
		\begin{pmatrix}
			  0&   1&   -9&   -9&  3&  3\\
			  1&    0&   -8&   -8&  4&  4\\
			  0&   0&   -6&   -6&  6&  6\\
			  0&   0&   -12&  0&   0&  12
		\end{pmatrix}\, .
	\end{equation}
		We define a $\mathbb{Z}_2$ involution $\tilde{\mathcal{I}}:=L\circ \phi_{[t]}$, via
	\begin{equation}\label{eq:example3_involution}
		t=\frac{1}{2}\begin{pmatrix}
			1\\
			0\\
			0\\
			0
		\end{pmatrix}\, ,\quad L=\begin{pmatrix}
			1&  0& -2&  0\\
			0&  1& -2&  0\\
			0&  0& -1&  0\\
			0&  0&  0& -1
		\end{pmatrix}\, .
	\end{equation} 
	Symmetrizing the heights of the Delaunay triangulation of $\Delta^\circ$ as in \S\ref{sec:invariant_toric_fans} leads to a fine, regular and star subdivision $\mathcal{T}$ defining a toric fan with a single non-simplicial cone $\sigma^*$.  Let $V$ be the corresponding (singular) toric variety. The generic CY hypersurface does not intersect the singular curve in $V$, and is thus smooth. We have
	\begin{equation}
		h^{1,1}(V)=243\, .
	\end{equation}
	By computing the induced action of $L$ on $H^2(V)$ we find
	\begin{equation}
		h^{1,1}_+(V,\tilde{\mathcal{I}})=123\, ,\quad h^{1,1}_-(V,\tilde{\mathcal{I}})=120\, .
	\end{equation}
	The $L$-invariant sublattice of $L$ is two-dimensional, and the diagonal toric variety $V_L$ is thus a toric surface, with edges defined by three vertices given by the invariant points
	\begin{equation}
		\begin{pmatrix}
			p_1&p_2&p_{146}
		\end{pmatrix}=\begin{pmatrix}
			0&  1& -3\\
			1&  0& -2\\
			0&  0&  0\\
			0&  0&  0
		\end{pmatrix}\, 
	\end{equation}
	 defining a $2d$ reflexive sub-polytope. We have $V_L\simeq \mathbb{P}^2_{[3,2,1]}$.
	
	In homogeneous coordinates the involution \eqref{eq:example3_involution} can be represented as
	\begin{equation}
		x_2\mapsto -x_2\, ,\quad x_3\leftrightarrow x_6\, ,\quad 
		x_4\leftrightarrow x_5\, ,\ldots\, ,
	\end{equation}
	where we have omitted all pairs of homogeneous coordinates that are permuted, but are not associated with vertices of $\Delta^\circ$. We define an involution on $X$ by setting $\lambda_f=1$. Then, the generic $\mathbb{Z}_2$ covariant anti-canonical polynomial reads
	\begin{equation}
		f=\psi_0 x_1\cdots x_6+\psi_1 x_2^3+\psi_2 x_2 x_3^4x_4^4x_5^4x_6^4 +\psi_3 \left(x_3^{12}x_4^{12}-x_5^{12}x_6^{12}\right)+
		\psi_4\left(x_3^{12}x_6^{12}-x_4^{12}x_5^{12}\right)\, ,
	\end{equation}
	where we have suppressed the dependence of the monomials on all homogeneous coordinates not associated with vertices of $\Delta^\circ$. The polynomial $f$ defines a smooth orientifold invariant Calabi-Yau hypersurface in $X$.
	
	Next, we compute the fixed point set in $V$. We have that
	\begin{equation}
		t+\frac{1}{2}p_2\in N\, ,\quad t+\frac{1}{2}p_{146}\in N\, ,
	\end{equation}
	and therefore the fixed point set $\mathcal{F}_{\tilde{\mathcal{I}}}\subset V$ contains the toric curves $\mathcal{F}_{\tilde{\mathcal{I}}}(p_2,0)=\hat{D}_{p_2}\cap V_L$ and $\mathcal{F}_{\tilde{\mathcal{I}}}(p_{146},0)=\hat{D}_{p_{146}}\cap V_L$ as irreducible components. The full fixed point set contains further copies of these curves associated with non-trivial elements $\nu\in H^L_-$, as we will see now. The lattice $P^L_-(N)$ is generated by
	\begin{equation}
		\ell_1=\begin{pmatrix}
			0 & 0 & 0 & 1
		\end{pmatrix}\, ,\quad \ell_2=\begin{pmatrix}
		1 & 1 & 1 & 0
	\end{pmatrix}\, ,
	\end{equation}
	and therefore, according to \eqref{eq:equivalencerelationsigns}, there are four  fixed curves $\mathcal{F}_{\tilde{\mathcal{I}}}(p_2,\nu_{s_1,s_2})$ and four fixed curves $\mathcal{F}_{\tilde{\mathcal{I}}}(p_{146},\nu_{s_1,s_2})$ associated with the vectors
	\begin{equation}
		\nu_{s_1,s_2}=s_1\ell_1+s_2\ell_2\, ,\quad \text{with} \quad s_i\in \{0,1\}\, .
	\end{equation}
	The resulting fixed point set in $X$ is a set of isolated points, and their number is computed using intersection theory. We have
	\begin{equation}
		\#\left(\mathcal{F}_{\mathcal{I}}(p_2,\nu_{s_1,s_2})\right)=3\, ,\quad \#\left(\mathcal{F}_{\mathcal{I}}(p_2,\nu_{s_1,s_2})\right)=1\, .
	\end{equation}
	Thus, in total we find
	\begin{equation}
		N_{\text{O3}}=4\times (3+1)=16
	\end{equation}
	isolated fixed points, in other words $16$ O3 planes. We have $h^{1,1}_-(X,\mathcal{I})=h^{1,1}_-(V,\tilde{\mathcal{I}})=120$ and we compute from the fixed point theorem that
	\begin{equation}
		h^{2,1}_-(X,\mathcal{I})=3=h^{2,1}(X)\, .
	\end{equation}
	The D3-tadpole is $Q_{D3}=\frac{\chi(\mathcal{F})}{4}=4$.

	Therefore, complex structure moduli space remains entirely unobstructed by the orientifold projection. In particular the generic $\mathbb{Z}_2$ invariant Calabi-Yau is indeed smooth. Just as in \ref{sec:491} the polytope $\Delta^\circ$ is trilayer. Thus, there exists another distinct smooth orientifold that is completely analogous to the one described in \ref{sec:491}, but no further that satisfy our smoothness criteria \S\ref{sec:smoothness}.

	\section{Conclusions}\label{sec:conclusions}
	In this work we have developed the geometric tools that allow the systematic enumeration and exploration of Calabi-Yau orientifolds that arise via orientifolding Calabi-Yau threefold hypersurfaces from the Kreuzer-Skarke list \cite{Kreuzer:2000xy}. In particular, as demonstrated in the example section \S\ref{sec:examples}, our methods enable the treatment of even the largest Hodge numbers.
	
	It will be interesting to apply our methods to continue to explore the string axiverse \cite{Arvanitaki:2009fg,Conlon:2006tq,Svrcek:2006yi}, specifically at large number of axions as in \cite{Demirtas:2021gsq}, but with regards to the spectrum of axions that arises from dimensionally reducing the ten-dimensional two-forms of type IIB string theory, cf. \cite{McAllister:2008hb,Hebecker:2015tzo,Hebecker:2018yxs,Carta:2020ohw,Carta:2021uwv,Cicoli:2021gss,Cicoli:2021tzt,Carta:2022web}.
	
	We note that our methods are straightforwardly implemented with algorithms that run in polynomial time in $h^{1,1}$. This allows enumerating all orientifolds that arise from polytopes with small $h^{1,1}$ in negligible time, while enumerating orientifolds of Calabi-Yau hypersurfaces at the largest values of $h^{1,1}$ in mere seconds on a laptop. We plan to make our algorithms available via a future release of $\mathtt{CYTools}$.

	\section*{Acknowledgments}
	We would like to thank Manki Kim for collaboration in early stages of this project, and Andres Rios-Tascon for patiently explaining various aspects of the triangulation of polytopes to us. We thank Naomi Gendler, Manki Kim, Liam McAllister, and Andres-Rios Tascon for useful comments on a draft. Furthermore, we thank Mehmet Demirtas, Geoffrey Fatin, Naomi Gendler, Manki Kim, Nate MacFadden, Liam McAllister, Richard Nally, Andres Rios-Tascon, Andreas Schachner and Mike Stillman for many helpful discussions. This research was supported in part by NSF grant PHY-2014071.
\bibliography{biblio}
\bibliographystyle{JHEP}
\end{document}